\documentclass[cernpreprint, atlasdraft=false, texlive=2023, UKenglish, orcidlogo, texmf, orcidlogo]{atlasdoc}
 
\usepackage{atlaspackage}
\usepackage{multirow}
\usepackage{atlasbiblatex}
 
\usepackage{graphics}
\usepackage{caption}
\usepackage{diagbox}
\usepackage{amsmath}
\usepackage{graphicx}
\usepackage{multirow}
\usepackage{upgreek}
\usepackage{dcolumn}
 
\usepackage{booktabs}
\usepackage{siunitx}
\usepackage{siunitx}
\usepackage{tikz}
 
\usepackage{tablefootnote}
 
\usepackage{xspace}

\usepackage{atlasphysics}

\graphicspath{{logos/}{figures/}}

\AtlasTitle{\vspace{-1.2cm}Search for resonant $WZ$ production in the fully leptonic final state in proton--proton collisions at $\mathbf{\sqrt{s} = 13}$~TeV with the ATLAS detector}
\AtlasJournalRef{Eur. Phys. J. C 83 (2023) 633}
\AtlasDOI{10.1140/epjc/s10052-023-11437-7}
 
\AtlasVersion{3.0}
 
\AtlasAbstract{
A search for a $WZ$ resonance, in the fully  leptonic final state  (electrons or muons), is performed using 139~fb$^{-1}$ of data collected at a centre-of-mass energy of 13~TeV by the ATLAS detector at the Large Hadron Collider.  The results are interpreted in terms of a singly charged Higgs boson of the Georgi--Machacek model, produced by $WZ$ fusion, and of a Heavy Vector Triplet, with the resonance produced by $WZ$ fusion or the Drell--Yan process.  No significant excess over the Standard
Model prediction is observed and limits are set on the production cross-section times branching ratio as a function of the resonance mass for these processes.
}

\PreprintIdNumber{CERN-EP-2022-093}

\AtlasJournal{EPJC}

\hypersetup{pdftitle={ATLAS document},pdfauthor={The ATLAS Collaboration}}

\begin{document}
 
\maketitle
 
\tableofcontents
 
\section{Introduction}

\label{sec:introduction}
Searches for diboson resonances provide an essential test of theories of electroweak symmetry breaking beyond the Standard Model (BSM): new charged scalar diboson resonances arise in various models with an extended Higgs sector~\cite{LHCHiggsCrossSectionWorkingGroup:2016ypw,Ivanov:2017dad,Steggemann:2020egv} and vector resonances are predicted in various
BSM scenarios~\cite{Agashe:2004rs,Giudice:2007fh,Foadi:2007ue,Randall:1999ee,Csaki:2003dt,Csaki:2015hcd}.
In this article, a search for a  \textit{ WZ} resonance produced via either the Drell--Yan process or vector-boson fusion (VBF) is conducted in the fully leptonic decay channel $ \ell \nu \ell \ell$ ($\ell  = e ~\mathrm{or} ~\mu$) in proton–proton ($pp$) collisions. The $pp$ collision data, with an integrated luminosity of 139 fb$^{-1}$, were collected by the ATLAS detector~\cite{PERF-2007-01} at the Large Hadron Collider (LHC) at a centre-of-mass energy of $\sqrt{s} = 13$ TeV.

In the Minimal Supersymmetric Standard Model the tree-level coupling of the charged Higgs boson to \textit{ WZ} is loop-induced~\cite{Djouadi:2005gj},
and therefore strongly suppressed compared to fermionic couplings. Tree-level coupling to massive vector bosons, is, however, present in extensions of the Standard Model (SM) with higher-isospin scalar fields~\cite{Mohapatra:1974hk,Arkani-Hamed:2002ikv,Han:2003wu}.
In this article, the Georgi--Machacek (GM) model~\cite{Georgi:1985nv,Chanowitz:1985ug} is used as a benchmark. Because it preserves custodial symmetry at tree level, it is not strongly constrained~\cite{Hartling:2014aga}.
The GM model extends the Higgs sector of the Standard Model by including one real and one complex triplet.
A parameter, $\sin \theta_H $, representing the mixing of the vacuum expectation values, determines the contribution of the triplets to the masses of the $W$ and $Z$ bosons.
The  physical scalar states are organized into different custodial multiplets: a  fiveplet $(H_5^{++},H_5^+,H_5^0,H_5^-,H_5^{--})$ that is fermiophobic but couples to $W$ and $Z$ bosons,  a  triplet,
and two singlets, one of which is identified as the observed $125~\GeV$ Higgs boson with SM properties. Single production of $H^{\pm}_5$ occurs by vector-boson fusion and, in this analysis, the assumption that the triplet states are heavier than the fiveplet scalars implies that it can only decay to $W^\pm Z$. The cross-section is proportional to $\sin^2 \theta_H$. The singly charged members of this fiveplet are the object of the present search in the VBF channel.
 
Parameterized Lagrangians~\cite{deBlas:2012qp,Pappadopulo:2014qza,Greco:2014aza} incorporating a Heavy Vector Triplet (HVT) allow the results of searches for vector resonances to be interpreted in a generic way.
In this article, a simplified phenomenological Lagrangian~\cite{Pappadopulo:2014qza} is used.
The new heavy vector resonance, $W'$, couples to the Higgs field and longitudinally polarized SM gauge bosons by virtue of the equivalence theorem~\cite{Chanowitz:1985hj}, and this coupling is parameterized by the product of two parameters $g_Vc_H$. It couples
to the fermions via the combination $(g^2/g_V)c_F$, where $g$ is the SM SU(2) gauge coupling. The parameter $g_V$ represents the typical strength of the vector-boson interaction, while the parameters $c_H$ and $c_F$ are expected to be of the order of unity in most models. The vector-boson scattering process, $pp \to W' jj \to WZ jj$,  is only sensitive to the gauge boson coupling and, in this case, the benchmark model used to interpret the results assumes no coupling of the heavy vector resonance to fermions.
 
\begin{figure}[htbp]
 
\includegraphics[width= 0.95\textwidth]{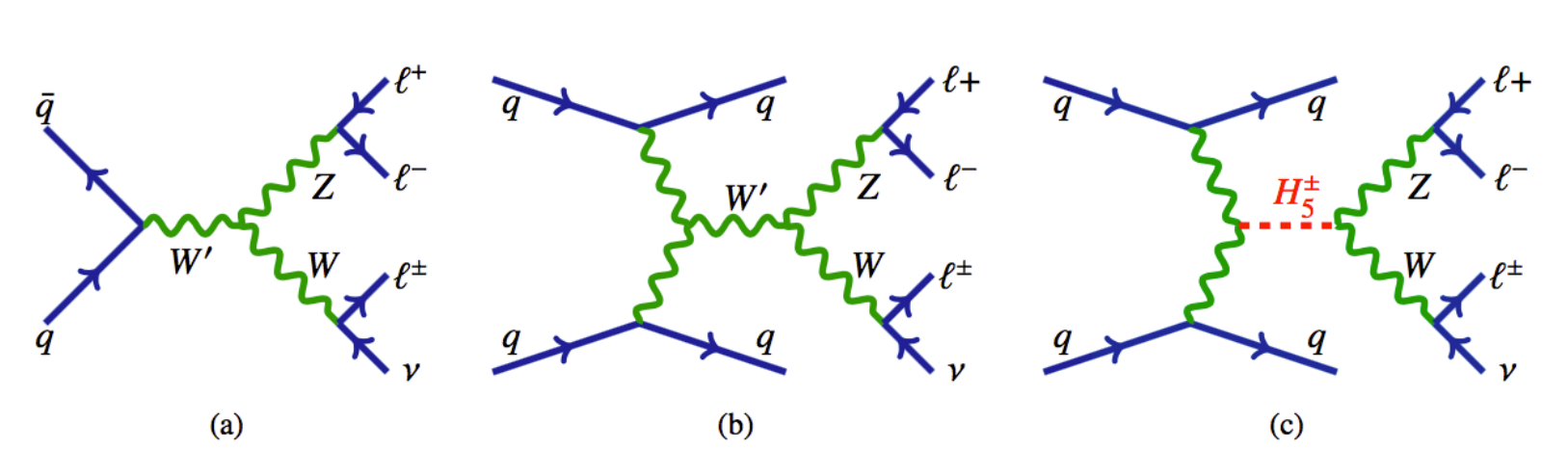}
\label{fig:Feynman}
 
\caption{Representative Feynman diagrams for heavy resonance production and decay to \textit{ WZ} bosons (a) HVT $W'$ production via Drell-Yan, (b) HVT $W'$ production via Vector Boson Fusion and (c) GM $H^{\pm}_5$ production via Vector Boson Fusion. The subsequent decays to the $\ell^{+}\ell^{-}\ell^{\pm}\nu$ are also shown. }
\label{FD}
 
\end{figure}

In nearly all of the parameter space explored in the present analysis and for both  benchmark models, the intrinsic width of the resonance is below 4\% , which is less than the experimental resolution.
Results are provided for the VBF and Drell--Yan production modes separately for the HVT process, neglecting possible signal leakage between them since the VBF contribution is quite small relative to Drell-Yan, always below 1\%, and since the VBF benchmark model considered here assumes no coupling to fermions.
Representative Feynman diagrams for the production and decay of the heavy resonances searched for in the present analysis are shown in Figure~\ref{FD}.

Searches for a $W'$ in an extended gauge model, decaying to \textit{ WZ} in the fully leptonic mode, at $\sqrt{{s}}=8~\TeV$ with 20~fb$^{-1}$ of data have been performed by the ATLAS~\cite{EXOT-2013-07} and CMS~\cite{CMS-EXO-12-025} Collaborations. The present analysis extends searches for resonant \textit{ WZ} production, performed by ATLAS in
Run 2 of the LHC using $pp$ collision data at
$\sqrt{{s}}=13~\TeV$~\cite{EXOT-2016-11},  with
36~fb$^{-1}$ of integrated luminosity.
 
Exclusion limits from searches of diboson resonances with different final states are summarized in Refs.~\cite{ATLAS-summary,ATL-PHYS-PUB-2021-018,CMS-summary}.
The results  from searches for heavy $VV$ and $VH$ vector resonances ($V = W$ or $Z$) and their combination, based on Run 1 data  and on Run 2 data in the fully hadronic ($qqqq$), semileptonic ($\ell\nu qq, \ell\ell qq, \nu\nu qq$), and fully leptonic ($\ell\ell\ell\ell$, $\ell\nu\ell\ell$, $\ell\ell\nu\nu$) final states are given in Refs.~\cite{EXOT-2014-18,EXOT-2016-01,HDBS-2018-10,CMS-B2G-16-007,CMS-B2G-18-006,CMS-B2G-19-002}.
The various decay channels generally differ in sensitivity in different mass regions.
The fully leptonic channel is found to be more sensitive to resonances with mass below $\sim$~1 TeV because of the low background, in spite of the low branching ratio.
For the VBF process, the present analysis aims to complement previous explorations of the HVT phase space since other channels are mostly insensitive when the coupling of the heavy vectors to fermions is close to zero.
 
Limits on the GM model have also been set, based on an analysis of opposite-charge \textit{ WW} production by ATLAS~\cite{HIGG-2016-31}, using data at $\sqrt{s}=13~\TeV$.
Searches~\cite{CMS-SMP-17-004,CMS-HIG-20-017} by the CMS Collaboration for a singly-charged  and a doubly-charged Higgs boson,  produced via VBF and decaying respectively into \textit{ WZ} and \textit{ WW} in the fully leptonic mode, using
an integrated luminosity of 137~fb$^{-1}$, have yielded limits on the coupling parameter of the GM model, assuming degenerate masses of $H_5^{\pm}$  and $H_5^{\pm\pm}$.
Upper bounds at 95\% confidence level (CL) on $\sin\theta_H$   vary between $\sim 0.2$ and 0.55 in the mass range 200-2000 GeV. In the present analysis, in addition to the larger data set, several improvements relative to the previously published analysis~\cite{EXOT-2016-11} have been implemented, most notably the implementation of multivariate techniques for the VBF signal selection.


\section{The ATLAS Detector}

\label{sec:detector}
\newcommand{\AtlasCoordFootnote}{
ATLAS uses a right-handed coordinate system with its origin at the nominal interaction point (IP)
in the centre of the detector and the $z$-axis along the beam pipe.
The $x$-axis points from the IP to the centre of the LHC ring,
and the $y$-axis points upwards.
Cylindrical coordinates $(r,\phi)$ are used in the transverse plane,
$\phi$ being the azimuthal angle around the $z$-axis.
The pseudorapidity is defined in terms of the polar angle $\theta$ as $\eta = -\ln\tan(\theta/2)$.
Angular distance is measured in units of $\Delta R \equiv \sqrt{(\Delta\eta)^{2} + (\Delta\phi)^{2}}$.}
 
The ATLAS detector~\cite{PERF-2007-01} has a
cylindrical geometry with a nearly $4\pi$ coverage in
solid angle\footnote{\AtlasCoordFootnote}.
The inner detector (ID), consisting of silicon pixel, silicon microstrip and transition radiation detectors, is surrounded by a thin superconducting solenoid
providing a \SI{2}{\tesla} axial magnetic field. It allows precise reconstruction of tracks from charged particles and measurement of their momenta up to a pseudorapidity of $|\eta| = 2.5$.
High-granularity lead/liquid-argon (LAr) sampling electromagnetic and steel/scintillator-tile hadron calorimeters, at larger radius,  provide energy measurements in the central pseudorapidity range $|\eta| < 1.7$. In the endcap and forward regions, LAr calorimeters
for both the electromagnetic and hadronic energy measurements extend the region of angular acceptance up to $|\eta| = 4.9$.
Outside the calorimeters, the muon spectrometer incorporates multiple layers of trigger and tracking chambers in a magnetic field produced by a system of superconducting toroid magnets, enabling an independent precise measurement of muon track momenta for $|\eta|<2.7$.
The ATLAS trigger system consists of a hardware-based level-1 trigger followed by a software-based high-level trigger~\cite{TRIG-2016-01}. An extensive software suite~\cite{ATL-SOFT-PUB-2021-001} is used in the reconstruction and analysis of real and simulated data, in detector operations, and in the trigger and data acquisition systems of the experiment.


\section{Data and Monte Carlo Samples}

\label{sec:Data}
 
\label{Data}
 
The data used were collected from 2015 to 2018 with the ATLAS detector from $pp$ collisions at a centre-of-mass energy of $13~\TeV$ at the LHC, and initially selected by requiring  that a set of quality criteria for detector and data conditions be satisfied~\cite{DAPR-2018-01}.
 
Events were required to pass combinations of  single-electron or single-muon triggers~\cite{TRIG-2018-05,TRIG-2018-01}.  The transverse momentum (\pT) thresholds of the leptons in 2015 were 24~GeV for electrons and 20~GeV for muons, with both satisfying a loose isolation requirement based only on ID track information. Due to the higher instantaneous luminosity in 2016--2018 the trigger threshold was increased to 26~GeV for both the electrons and muons, and tighter isolation requirements were applied.
An additional electron (muon) trigger with a \pT threshold of 60 (50)~GeV and no isolation requirement, and a single-electron trigger requiring \pT $>$~120~GeV with less restrictive electron identification criteria, were used to increase the selection efficiency, which reached almost 100\%, relative to the offline selections~\cite{TRIG-2018-05,TRIG-2018-01}.  With these conditions, the integrated luminosity used in this analysis is 139~fb$^{-1}$.
 
Simulated signal events and background processes with prompt leptons were used to model the benchmark physics processes and optimize the selection cuts. They were produced by Monte Carlo (MC) generators with the detector response modelled by
the $\GEANT$ toolkit~\cite{Agostinelli:2002hh,SOFT-2010-01} integrated into the ATLAS simulation infrastructure. For some samples, the calorimeter response is obtained from  a fast parameterized detector simulation~\cite{ATL-PHYS-PUB-2010-013}, instead of full simulation by $\GEANT$. The effect of multiple interactions in the same and neighbouring bunch
crossings (\pileup) was modelled by overlaying the simulated hard-scattering event with inelastic $pp$ events generated with \PYTHIA[8.186]~\cite{Sjostrand:2007gs}
using the \verb+NNPDF2.3LO+ set of parton distribution functions (PDF)~\cite{Ball:2012cx} and the A3 set of tuned parameters (tune)~\cite{ATL-PHYS-PUB-2016-017}. The distribution of the number of pile-up events reproduces the bunch structure and the average number of interactions per bunch crossing in the various run periods. A pileup weight is defined, which is applied to Monte Carlo to correct for the difference between the distribution of average number of interactions used to produce the sample and that measured for the recorded data. For all samples, except
those generated with \textsc{Sherpa}~\cite{Gleisberg:2008ta}, the \EVTGEN[1.2.0] program~\cite{Lange:2001uf} was used to simulate the properties of
the $b$- and $c$-hadron decays.
 
The GM VBF benchmark signal samples,  $pp \to H_5^\pm jj \rightarrow W^\pm Z jj \to \ell^\pm \nu \ell^+\ell^- jj$, vetoing $W$ or $Z$ bosons in the s-channel, were produced with  \textsc{MadGraph} 2.7.2~\cite{Alwall:2014hca}
at next-to-leading order (NLO) in QCD~\cite{Degrande:2015xnm,LHCHiggsCrossSectionWorkingGroup:2016ypw};
the generator is referred to as \MADGRAPH hereafter.
The signal simulation is produced for the mass range 200~GeV to 1~TeV in the $H_5$-plane defined in Refs.~\cite{Logan:2017jpr,LHCHiggsCrossSectionWorkingGroup:2016ypw}, using the tool \textsc{GMCALC}~\cite{Hartling:2014xma}. The parameter $\sin \theta_H$ was set to 0.5 for masses up to 800~GeV and 0.25 for higher masses to be compatible with present constraints~\cite{Logan:2017jpr}. The matrix element calculation employed the \texttt{NNPDF3.0NLO}~\cite{Ball:2014uwa} set of PDFs. Events were interfaced to \textsc{Pythia}~8.186 for the modelling of the parton shower, hadronization, and underlying event, using the A14 tuning parameters~\cite{ATL-PHYS-PUB-2014-021} and with the dipole recoil shower scheme to prevent the generation of excess central jet radiation~\cite{Ballestrero:2018anz}. For these samples, a minimum \pT of $15~\GeV$ ($10~\GeV$) for the jets (leptons) was required during event generation. The signal samples were produced in $25~\GeV$ mass steps up to $600~\GeV$ and $100~\GeV$ mass steps up to $1~\TeV$.

Two benchmark models of HVT production via the Drell--Yan process, $q q'\to W' \to WZ \to \ell\nu\ell\ell$, are used to interpret the results.
Model A is typical of weakly coupled vector resonances arising from an extension of the SM gauge group~\cite{Barger:1980ix} with an additional SU(2) symmetry, and the branching ratios to fermions and gauge bosons are comparable.
Model B is representative of an HVT produced in a strongly-coupled scenario, as in a Composite Higgs model~\cite{Contino:2011np} with suppressed fermionic couplings.
The parameter $g_V$ was set to 1 for Model A and to 3 for Model B. For both models,  $c_F$ is set to 1 and is assumed to be the same for all types of fermions.
The simulated signal samples for Model A were generated
at leading order (LO) in QCD with \textsc{MadGraph}~2.6.5 using the model file provided by the authors of Ref.~\cite{Pappadopulo:2014qza}. The parton-level simulated events were hadronized with \textsc{Pythia~8.186}, using the \verb+NNPDF23_lo_as_0130_qed+ PDF set and A14 tune. The signal samples were produced for vector resonances with masses ranging from $250~\GeV$ to $5~\TeV$, in steps of $25~\GeV$ below $600~\GeV$, $100~\GeV$ between $600~\GeV$ and $2~\TeV$, $200~\GeV$ from $2~\TeV$ to $3~\TeV$ and $500~\GeV$ above. For interpretation in terms of Model B, the Model A simulation is used and the cross-sections were simply scaled. This is justified since the intrinsic resonance width remains well below the experimental resolution and the angular distributions are the same for both models.
 
For the VBF production mode of heavy vector resonances, which is expected to have a low cross-section, the benchmark model used is also based on the HVT parameterization.
The coupling parameters $g_V$ and $c_H$ are set to 1 and all other couplings of the heavy triplet, including $c_F$, are set to 0  in order to maximize the VBF contribution.
The simulated signal samples were generated at LO in QCD with \textsc{MadGraph}~2.6.5 using the model file provided by the authors of Ref.~\cite{Pappadopulo:2014qza}. The parton-level simulated events were hadronized with \textsc{Pythia~8.186}, using the \verb+NNPDF23_lo_as_0130_qed+ PDF set and A14 tune. A dijet invariant mass of at least $150~\GeV$ is required in this case at  event generation. The simulation samples were generated for masses ranging from $300~\GeV$ to $2~\TeV$, in steps of $25~\GeV$ ($100~\GeV$) up to (beyond) $600~\GeV$.
 
The background sources include processes with two or more electroweak gauge bosons, namely $VV$ and $VVV$ ($V=Z$, $W$) as well as processes with top quarks, such as $t\bar{t}$, $t\bar{t}V$, and single top-quark, and processes with gauge bosons produced in association with jets or photons.
 
The dominant background for this search is the SM QCD mediated \textit{ WZ} process, referred to as \textit{ WZ}-QCD here. It includes processes up to order four in the electroweak coupling constant, $\alpha_{EW}$, and is modelled using \textsc{Sherpa}~2.2.2~\cite{Gleisberg:2008ta}.
The \textit{ WZ} sample includes up to one jet calculated at NLO in QCD, while second and third jets were calculated at LO in QCD and merged with the parton shower.
In order to estimate an uncertainty due to generator and parton shower modelling, an alternative NLO \textit{ WZ}-QCD sample was produced using \textsc{MadGraph} 2.6.5 with FxFx merging~\cite{Frederix:2012ps} of up to two extra jets, using the PDF set  \verb+NNPDF30_nlo_as_118+. The hadronization was performed with \textsc{Pythia~8.186} with the A14 tune.
A sample of the purely electroweak process $WZjj \to \ell\nu~\ell\ell~jj$, including processes of order six in $\alpha_{EW}$ (\textit{ WZ}-EWK), was generated separately with \textsc{MadGraph}~2.7.3 together with \textsc{Pythia~8.244} using the A14 tune~\cite{ATL-PHYS-PUB-2014-021} and the
\verb+NNPDF3.0NLO+\cite{Ball:2014uwa} PDF set. To estimate an uncertainty due to the parton shower modelling, a sample using the same \textsc{MadGraph}~2.7.3 matrix element but \textsc{Herwig~7.2.1} for the parton shower was produced. According to the SM, a small amount of interference occurs between electroweak and QCD \textit{ WZ} production.
This was modelled with \textsc{MadGraph}~2.7.3 \textsc{+ Pythia~8.244} using the A14 tune~\cite{ATL-PHYS-PUB-2014-021} and the \verb+NNPDF3.0NLO+\cite{Ball:2014uwa} PDF set and combined with the simulated \textit{ WZ}-EWK sample.
 
Samples of $q\bar q\to ZZ \to 4\ell$, $q\bar q\to ZZ \to \ell\ell~\nu\nu$
and triboson events were generated with \textsc{Sherpa}~2.2.2~\cite{Bothmann:2019yzt} using matrix elements at NLO accuracy in QCD for up to one additional parton and at LO accuracy for up to three additional parton emissions. The simulation  included off-shell effects and Higgs boson contributions. The purely electroweak process $q\bar q\to ZZjj \to 4\ell jj$ and the $gg\to ZZ$ process were also generated with \textsc{Sherpa}~2.2.2~\cite{Bothmann:2019yzt}. The LO-accurate matrix elements were matched to a parton shower based on Catani--Seymour dipole factorization~\cite{Gleisberg:2008fv,Schumann:2007mg} using the MEPS@LO
prescription~\cite{Hoeche:2011fd,Hoeche:2012yf,Catani:2001cc,Hoeche:2009rj}.
Samples were generated using the \verb+NNPDF3.0NNLO+ PDF set~\cite{Ball:2014uwa}, and \SHERPA parton-shower parameter values.

The $t\bar{t}V$ processes were modelled using the
\textsc{MadGraph}~2.3.3~\cite{Alwall:2014hca} generator at NLO in QCD with the
\verb+NNPDF3.0NLO+\cite{Ball:2014uwa} PDF set.
The events were interfaced to \PYTHIA[8.210]~\cite{Sjostrand:2014zea}
using the A14 tune and the
\verb+NNPDF2.3LO+~\cite{Ball:2014uwa} PDF set.
 
Finally, samples of SM backgrounds with at least one misidentified or non-prompt lepton, including $Z\gamma$, $W\gamma$, Drell--Yan $Z\to \ell \ell,~W\to\ell\nu$ as well as top-quark pairs and single top-quark have been generated to assist in estimating the fake/non-prompt lepton background.
Events with $V\gamma$ in the final state were simulated with the
\SHERPA[2.2.4]~\cite{Bothmann:2019yzt} generator. Matrix elements at LO
accuracy in QCD for up to three additional parton emissions were
matched and merged with the \SHERPA parton shower~\cite{Gleisberg:2008fv,Schumann:2007mg,Hoeche:2011fd,Hoeche:2012yf,Catani:2001cc,Hoeche:2009rj}. The samples were generated using the \verb+NNPDF3.0NNLO+ PDF set~\cite{Ball:2014uwa},
along with the dedicated set of tuned parton-shower parameters
developed by the \SHERPA authors. Drell--Yan $Z\to \ell \ell$ and $W\to\ell\nu$ were produced with \POWHEGBOX[v1] generator~\cite{Nason:2004rx,Frixione:2007vw,Alioli:2010xd,Alioli:2008gx} at NLO accuracy for the hard-scattering processes of $W$
and $Z$ boson production and decay in the electron, muon, and $\tau$-lepton
channels. The events were interfaced to \PYTHIA[8.186]~\cite{Sjostrand:2007gs}
for the modelling of the parton shower, hadronization, and underlying
event, with parameters set according to the \AZNLO
tune~\cite{STDM-2012-23}. The \verb+CT10NLO+ PDF set~\cite{Lai:2010vv} was used
for the hard-scattering processes, whereas the \CTEQ[6L1] PDF
set~\cite{Pumplin:2002vw} was used for the parton shower. The effect of
QED final-state radiation was simulated with \PHOTOSpp[3.52]~\cite{Golonka:2005pn,Davidson:2010ew}.  For top-quark pairs and single top-quark productions the \POWHEGBOX[v2]~\cite{Alioli:2009je,Nason:2004rx,Frixione:2007vw,Alioli:2010xd} generator was used at NLO with the \verb+NNPDF3.0NLO+~\cite{Ball:2014uwa} PDF set.
The events were interfaced with \PYTHIA[8.230]~\cite{Sjostrand:2014zea} using the A14 tune~\cite{ATL-PHYS-PUB-2014-021} and the
\verb+NNPDF2.3LO+ PDF set.
 
SM backgrounds with Higgs bosons ($H, t\bar tH, VH$) contribute less than 0.1\% of the total background because of the low cross-section and the requirement of a well-reconstructed leptonically decaying $Z$ boson. These backgrounds were neglected.


\section{Object reconstruction and identification}

\label{sec:Objects}
 
Electron candidates are reconstructed from energy deposits in the electromagnetic calorimeter which are matched to a well-reconstructed ID track~\cite{EGAM-2018-01}. Only electrons with transverse energy $\et>$7~\GeV~and within the pseudorapidity range of $|\eta|<$2.47, excluding the barrel--endcap transition region 1.37$<|\eta|<$1.52 are considered.
Muons are identified either by matching muon spectrometer tracks with tracks in the ID or by using the calorimeter-based muon identification~\cite{PERF-2015-10,MUON-2018-03}. They are required to have transverse momentum $\pt>$5~\GeV~~($\pt>$15~\GeV~if calorimeter tagged) and pseudorapidity $|\eta|<$2.7.
 
Identification and isolation criteria, either `loose', `medium' or `tight' as described in Refs.~\cite{EGAM-2018-01,PERF-2015-10}, are applied to electron and muon candidates. Identification criteria are based on shower shapes and track parameters for electrons, and on track parameters for muons. The isolation criteria  use information about ID tracks and calorimeter energy deposits in a fixed cone of size $\Delta R =$~0.2 around each lepton. Four lepton categories are designed using the identification and isolation criteria: \emph{Baseline} electrons and muons are required to satisfy `loose' identification and isolation criteria (for muons with $\pt>$300~\GeV~ the dedicated `High $\pt$ identification' is required~\cite{MUON-2018-03}).
The \emph{Loose}, \emph{Tight Z} and \emph{Tight W} leptons are defined as subsets of the \emph{Baseline} lepton selection with $\pt>$25~\GeV. For the \emph{Tight Z} leptons the `medium' identification and `tight' isolation criteria are applied, while for \emph{Tight W} leptons the `tight' identification and `tight' isolation criteria are applied. The tighter identification and isolation requirement applied to the lepton of the $W$ candidate is motivated by lower background rates for leptons from the $Z$ candidate,  which are well constrained by the requirement on their invariant mass.

Electron and muon candidates are required to originate from the primary vertex. The primary vertex is defined, using tracks with $\pt > $500~MeV, as the vertex candidate with the highest $\sum \pt ^2$ of its associated tracks. The transverse impact parameter of the track ($d_{0}$) is calculated relative to the beam line. For all four lepton categories, the longitudinal impact parameter, $z_0$ (the difference between the value of $z$ at the point of the track where $d_0$ is defined and the longitudinal position of the primary vertex), is required to satisfy $|z_0 \cdot \mathrm{sin}\theta|<$0.5~mm, where $\theta$ is the polar angle of the track momentum at the reference point. Furthermore, for the \emph{Loose}, \emph{Tight Z} and \emph{Tight W} leptons the significance of the transverse impact parameter of the track, $|d_{0}/\sigma_{d_0}|$, where $\sigma_{d_0}$ stands for the resolution of $d_{0}$, must be smaller than 3.0 for muons and less than 5.0 for electrons.

Jets are based on particle-flow objects built from noise-suppressed positive-energy topological clusters of cells in the calorimeter and reconstructed tracks~\cite{PERF-2015-09}. The anti-$k_t$ algorithm~\cite{Cacciari:2008gp,Fastjet} with a radius parameter of $R=$~0.4 is used.
For jets, the main backgrounds are either beam-induced, due to proton collisions upstream of the interaction point, from cosmic-ray showers or highly coherent calorimeter noise.
These jets are considered `unclean' and nearly all are rejected by applying a set of quality criteria. Furthermore, to mitigate contamination from the \pileup,  a jet vertex tagger~\cite{PERF-2014-03,ATLAS-CONF-2014-018}, using information about tracks associated with the primary vertex and \pileup vertices, is applied to jets with $\pt < 60~\GeV$ and $|\eta|<2.4$. In the forward region, \pileup jet tagging that exploits jet
shapes and topological jet correlations in \pileup interactions is applied to  jets with $\pt < 120~\GeV$ and $2.5<|\eta|<4.5$~\cite{ATLAS-CONF-2014-018}.
 
The flavour of jets is determined using a deep-learning neural network, DL1r~\cite{FTAG-2018-01,ATL-PHYS-PUB-2021-004}. The DL1r $b$-tagging is based on distinctive features of $b$-hadron decays in terms of the impact parameters of the tracks and the displaced vertices reconstructed in the inner detector.
The $b$-tagging algorithm has an efficiency of 85\% in simulated $t\bar{t}$ events, a light-flavour jet misidentification probability of 3\% and a $c$-jet misidentification probability of about 33\%.
 
Two levels of jet selections are used: the \emph{Baseline} jets have $\pt > 30$~\GeV~and $| \eta |<4.5$, while for \emph{VBF jets}, which are a subset of \emph{Baseline} jets, the \pileup removal using the jet vertex tagger and a $b$-tagging veto are applied, as described above, since they are mostly forward jets and not initiated by b-quarks.

To avoid cases where the detector response to a single physical object is reconstructed as two different
final-state objects, an overlap-removal procedure is applied to the \emph{Baseline} selected leptons and jets. If two electrons share the same track then the lower-$\pt$ electron is discarded. Electrons that share the same track as a selected muon with a muon spectrometer track are also discarded; but in the case of a calorimeter-tagged muon, it is the muon which is rejected. A jet is removed if its separation from an electron satisfies $\Delta R <$~0.2; the electron is removed if the separation satisfies 0.2~$<\Delta R <$~0.4. For nearly collinear muons and jets, the jet is removed if it is separated from the muon by $\Delta R <$~0.2 and if it has less than three tracks, or if the energy and momentum differences between the muon and the jet are small; otherwise the muon is removed if the separation satisfies $\Delta R <$~0.4.

The missing transverse momentum, $\met$, in an event is calculated as the magnitude of the negative vectorial sum of the transverse momenta of all \emph{Baseline} selected and calibrated physics objects that can be matched to the
primary vertex. A component called the “soft term” is calculated from the residual tracks that originate from the primary vertex but are not associated with any other object and is added to the $\met$ calculation~\cite{PERF-2016-07}.


\section{Event selection}

\label{sec:Selection}
 
In this search all final states with three charged leptons ($e$ or $\mu$) and missing transverse momentum from \textit{WZ} leptonic decays are considered. The search begins with a \textit{WZ} baseline selection, and two selections are defined in order to build signal regions (SRs) targeting the Drell--Yan and VBF productions modes. A cut-based selection is used to build the Drell--Yan signal region, while for the VBF selection, an artificial neural network (ANN) was trained. An alternative, cut-based selection for the VBF is also presented in the Appendix. The invariant mass of the \textit{WZ} candidates, $m$(\textit{WZ}), built with the leptons and \met is used as the discriminating variable.  A summary of all the selections used to define the analysis signal regions (SRs) and control regions (CRs) can be found in Table~\ref{tab:Selection}.

\begin{table}[h]
\caption{Summary of the event selections for signal and control regions.  Definitions of some variables used in this table can be found in Section~\ref{sec:DYSelection} and ~\ref{sec:VBFSelection}.}
\centering
\begin{tabular}{l|m{6cm}|m{5cm}}
\hline
\hline
\multicolumn{3}{c}{ \textbf{Baseline \textit{\textbf{WZ}} selection }}\\[0.5ex]
\hline
\multicolumn{3}{c}{Event cleaning and primary vertex} \\
\multicolumn{3}{c}{Single-electron or single-muon trigger} \\
\multicolumn{3}{c}{Exactly 3 \textit{Loose} leptons ($e$ or $\mu$) with $\pt>$25~$\GeV$ ($\pt>$27~$\GeV$ for the trigger-matched lepton)}\\
\multicolumn{3}{c}{\textit{ZZ} veto: veto events with additional \textit{Baseline} leptons}\\
\multicolumn{3}{c}{$Z$ candidate: A \textit{Tight Z} same-flavour-opposite-sign lepton pair with $|m_{\ell\ell}-m_Z|<20$~\GeV} \\
\multicolumn{3}{c}{$W$ candidate: \textit{Tight W}  lepton requirements on 'non-$Z$ leptons' and \met $>$ 25~\GeV} \\[0.5ex]
\hline
 
Selection  &  \textbf{Drell-Yan} & \textbf{VBF} \\[0.5ex]
 
\hline
\multirow{4}{*}{Signal region}   & $ \pt(V)/m(\textit{WZ})$ > 0.35   & At least 2 \emph{VBF jets}  \\
 
&  & $m_{\textit{jj}}$ > 100~\GeV~ \\
&   & Veto events with $b$-tagged jets \\
&   & ANN Output > 0.82   \\[0.5ex]
\hline
\multirow{3}{*}{\textit{WZ}-QCD control region} &  $ (\pt(W)/m(\textit{WZ}) \leq$ 0.35 or  &  At least 2 \emph{VBF jets}\\
&  $ \pt (Z)/m(\textit{WZ})\leq$ 0.35)   & $m_{\textit{ jj}}$ > 500~\GeV~ \\
& $ \pt(V)/m(\textit{WZ})$ > 0.1                                          & Veto events with $b$-tagged jets \\
&                                              & ANN Output < 0.82 \\[0.5ex]
\hline
\multirow{2}{*}{\textit{ZZ} control region}  & Additional \textit{Baseline} lepton  & Additional \textit{Baseline} lepton  \\
& No \met requirement                  & No \met requirement\\
&                                      & At least 2 \emph{VBF jets}\\[1ex]
 
\hline
\hline
\end{tabular}
\label{tab:Selection}
\end{table}
 
\subsection{ Baseline selection of \textit{WZ} events}
\label{subsec:preselection}
The baseline selection is a set of event criteria applied to data and all simulated samples before  defining more specific analysis regions.
First, there is a requirement of good quality for the recorded events, based on the working conditions of all subdetectors. Events are vetoed if they have one or more unclean jets. All events
are required to contain a primary vertex with at least two associated tracks.
 
Events are required to contain exactly three leptons meeting the \emph{Loose} selection criteria defined in Section~\ref{sec:Objects}. In order to reduce the \textit{ZZ} background, events with four or more leptons meeting the \emph{Baseline} criteria are vetoed. To ensure that the trigger efficiency is well
determined, at least one of the three candidate leptons must be trigger-matched and is required to have \pt$ > $27~\GeV.
A $Z$ candidate must be present. It is defined by two leptons of the same flavour and opposite charge with an invariant mass that is consistent with the Z boson pole mass ($m_Z$): $|m_{\ell \ell} - m_Z| < 20$ ~\GeV.  If there is more than one pair of leptons that can form a $Z$ candidate, the one with invariant mass closest to the $Z$ boson pole mass is chosen. The third lepton is then taken as the $W$ boson lepton candidate.
The leptons assigned to the $W$ and $Z$ candidates are then required to satisfy the \emph{Tight} $W$ or \emph{Tight} $Z$ selection criteria defined in Section~\ref{sec:Objects}. Finally, the missing transverse momentum in the event is required to be greater than 25~\GeV.
 
To reconstruct the four-vector of the $W$ boson,  the \MET of the event is assumed to be due to the neutrino. The longitudinal component $p_z(\nu)$ of the neutrino momentum is estimated by constraining the invariant mass of the $\ell\nu$ system to be the pole mass of the $W$ boson, where the charged lepton is the one assigned to the $W$ candidate. A quadratic equation leads to two solutions.  If they
are real, the one with the smaller magnitude of $|p_z(\nu)|$ is chosen, otherwise, the real part is chosen. The choice of the solution was optimized using truth information. The invariant mass of the \textit{WZ} system is then calculated.

\subsection{Drell--Yan process selection}
\label{sec:DYSelection}
\begin{figure}[!htbp]
\begin{center}
\includegraphics[width=.7\textwidth]{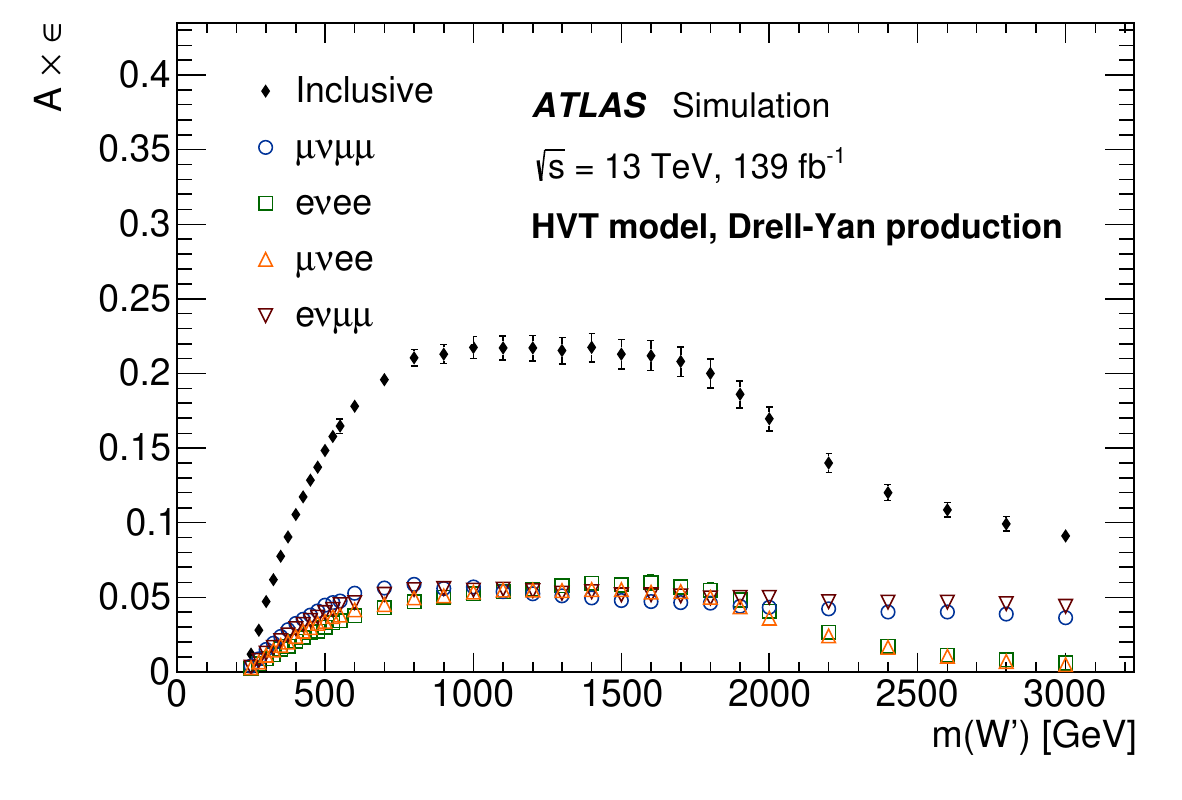}
\end{center}
\caption{The acceptance (A) times efficiency ($\epsilon$)  of the HVT $W'$ in the Drell--Yan signal region for different mass points and for the individual channels $\mu\nu\mu\mu$, $e\nu ee$, $\mu\nu ee$, $e\nu\mu\mu$, and the sum of all channels. The uncertainty includes both the  statistical and experimental systematic components. }
\label{fig:Results:DrellYannAcceptance}
\end{figure}

For a heavy resonance produced essentially at rest in the s-channel, it is expected that the selected $W$ and $Z$ bosons have transverse momenta close to 50\% of the resonance mass. A boson $\pt$ to resonance mass ratio variable is therefore defined as the ratio $\pt(V)/m(WZ)$ of the boson transverse momentum to the \textit{WZ} invariant mass. To reduce the contribution from  non-resonant \textit{WZ} production,  events passing the \textit{WZ} preselection are required to have a boson $\pt$ to resonance mass ratio greater than $0.35$ for both bosons. The combined detector acceptance and signal selection efficiency ($A\times\epsilon$) of the Drell-Yan HVT $W'$ selection, relative to the generated signal events, is shown in Figure~\ref{fig:Results:DrellYannAcceptance}. There, decays of $W$ and $Z$ bosons into all flavours of leptons are included at event generation. The $A\times\epsilon$ values decrease for resonance masses above approximately 2~TeV due to the collinearity of electrons from highly boosted $Z\xrightarrow[]{}ee$ decays, for which the lepton isolation is less efficient.
 
\subsection{Vector Boson Fusion process selection}
\label{sec:VBFSelection}
The VBF process ($pp \to W' j j \to WZ j j$) is characterized by the presence of two jets with a large rapidity gap resulting from quarks from which a vector boson has been radiated.
To select the signal events, an  artificial neural network with a binary classification task is used:  events are categorized as belonging either to a VBF process or to the background. The ANN training
is implemented using the {\textsc Keras} package~\cite{chollet2015keras} running on top of the  {\textsc TensorFlow} package~\cite{tensorflow2015-whitepaper}.
An ANN training region is defined by requiring events to have at least a pair of jets satisfying the \emph{VBF jets} selection, and
from those, the pair with the highest-$ \pt$ is required to have an invariant $m_{\textit{jj}}>$100~\GeV. The ANN is trained in this region with simulated $H^{\pm}_5$ events as signal, against the SM \textit{WZ}-EWK and \textit{WZ}-QCD events as background. The $H^{\pm}_5$ simulation is used for the training because the kinematic variables show very similar distributions for the GM and HVT benchmark signals and the training yielded similar results.

In order to minimize the statistical uncertainty, a 4-fold cross-validation technique was applied. A rectified linear unit, or a ReLU, was used as an activation function at each node~\cite{RELU}.
The space of hyperparameters was scanned and a final set was chosen to ensure optimal performance of the network.
The training was performed with 100 epochs, Nesterov's momentum of 0.7~\cite{pmlr-v28-sutskever13}, and two hidden layers of 45 neurons each.
To avoid overfitting, a regularization technique was employed. For each input sample, a hidden layer was randomly removed with a probability parameter (dropout) of 0.2, allowing for a noisy training process.
 
The distributions of the loss function and of accuracy vs epochs were monitored for the training and validations sets and no sign of overtraining was observed.

\begin{table}[h]
\caption{Variables used for ANN training. }
\centering
\begin{tabular}{c|c}       \hline
\hline
Training variable &  Definition \\[0.5ex] \hline
$m_{\textit{jj}}$        & Invariant mass of the two leading-\pt jets \\
$ \Delta\phi_{\textit{jj}}$ & Difference in $\phi$ of the two leading-\pt jets\\
$\eta_W, \eta_Z$  & Pseudorapidities of the reconstructed gauge bosons \\
$\eta_{j1}$ & Leading-\pt  jet pseudorapidity\\
$\zeta_{\mathrm{Lep}}$ & Event centrality\\
\MET & Missing transverse momentum \\
$H_{\textrm{T}}$ & Scalar \pt sum of the \emph{VBF jets} and the leptons from the $W\!Z$ decay \\[1ex]
\hline
\hline
\end{tabular}
\label{tab:MVAvariables}
\end{table}

The input variables\footnote{
The "event centrality" is a measure of the smaller pseudorapidity difference between the most forward jet and the most forward lepton in either hemisphere. It is defined as:
\begin{equation}
\nonumber
\zeta_{\mathrm{Lep}} = \min \left\{ \left[\min (\eta_{\ell_1}, \eta_{\ell_2},\eta_{\ell_3}) - \min (\eta_{j_1}, \eta_{j_2})], [\max (\eta_{j_1}, \eta_{j_2}) - \max (\eta_{\ell_1}, \eta_{\ell_2},\eta_{\ell_3})\right] \right\},
\label{eqn:centrality}
\end{equation}
with ${\ell_1}$, ${\ell_2}$, ${\ell_3}$ being the three leptons from the \textit{WZ} decay and $j_1$, $j_2$ the leading-\pt and subleading-\pt VBF jets.
}
used for the ANN optimization are listed in Table~\ref{tab:MVAvariables}.
These were chosen on the basis of their impact in the training and such that
highly correlated variables are not used simultaneously. The loss in expected significance when adding or exchanging some of the variables was evaluated for each set of variables until the optimal set was found.

All the mass samples of simulated $H^{\pm}_5$ GM events are used simultaneously to define the "signal" for the training.
After training,  the threshold for the ANN output score is chosen in such a way that it maximizes the significance for the lowest mass point (200~\GeV).
The advantage of this approach is that it greatly reduces the training effort and a single signal region can be used. It was verified that the alternative of using one ANN training per mass point does not significantly improve the performance.
The training is then applied to both GM and HVT Model samples. A minimum value of 0.82 on the ANN output maximizes the significance and is chosen to define the signal region.
After all selection cuts are applied the VBF signal region effectively starts at $m_{\textit{jj}}>500~\GeV$. This signal region was blinded until the background and its uncertainties in the control regions had been evaluated (Section~\ref{sec:Background}).

\begin{figure}[htbp]
\begin{center}
\includegraphics[width=.6\textwidth]{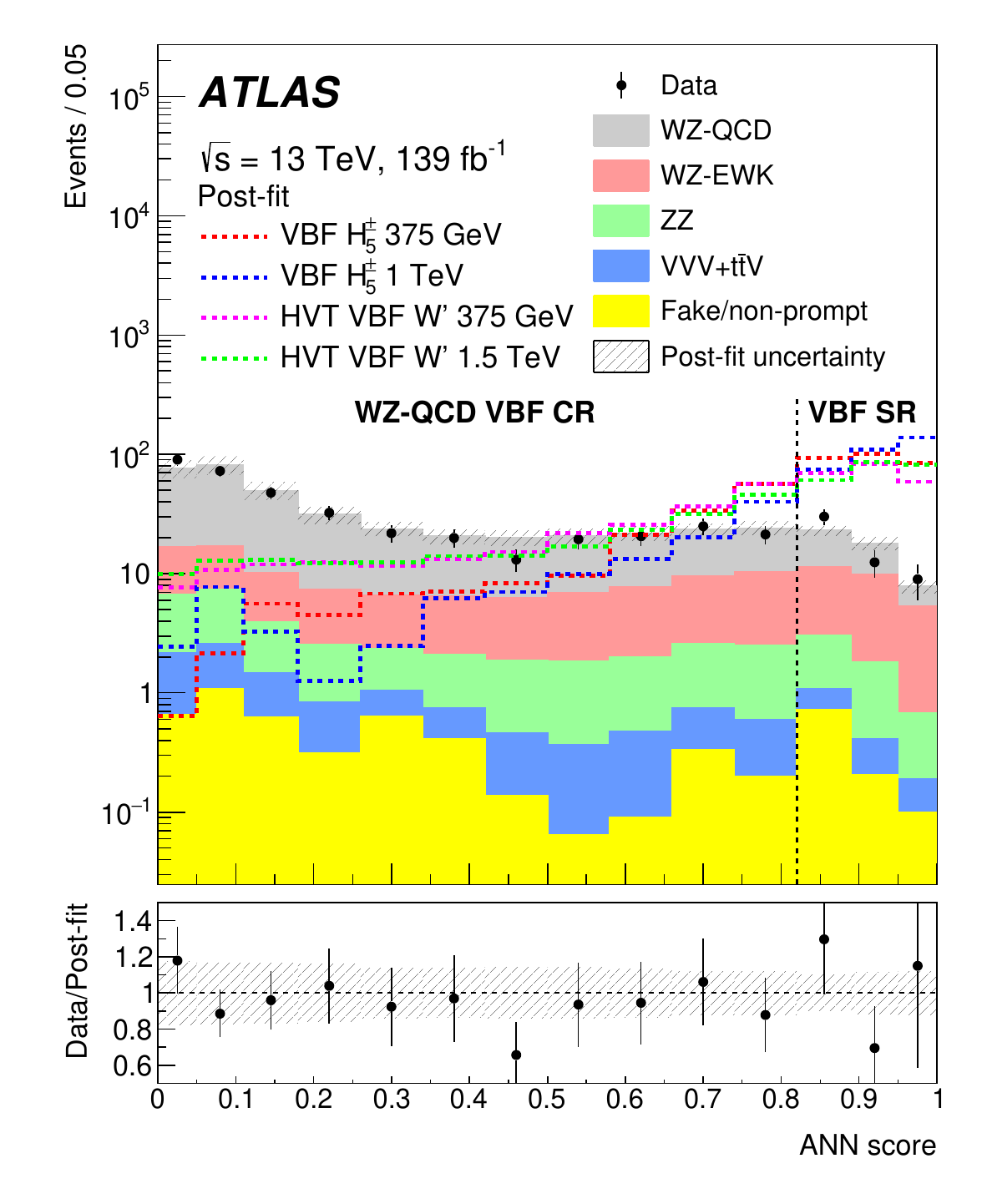}
\end{center}
\caption{Distribution of the ANN score in the \textit{WZ}-QCD VBF control region and the VBF signal region. The background predictions are obtained from a background-only simultaneous fit to the VBF signal region, the \textit{WZ}-QCD VBF and \textit{ZZ} VBF control regions as described in Section~\ref{sec:Results}. The uncertainty in the total background prediction, shown as a hashed area, combines statistical and systematic contributions. The distributions for the HVT VBF model $W'$ and GM $H_{5}^{\pm}$ simulations are shown normalized to the data integral.
The vertical dotted line shows the threshold value for the ANN output score used to define the VBF signal region. }
\label{fig:ANN_output}
\end{figure}

The distribution shapes and correlations of all input variables to the ANN were found to be well modelled by MC simulation in the \textit{WZ}-QCD control region (see Section~\ref{sec:Background} for definition). This is exemplified by the good description of the ANN output score distribution of data in the \textit{WZ}-QCD control region and VBF signal region shown in Figure~\ref{fig:ANN_output}.

\begin{figure}[!htbp]
\begin{center}
\includegraphics[width=.7\textwidth]{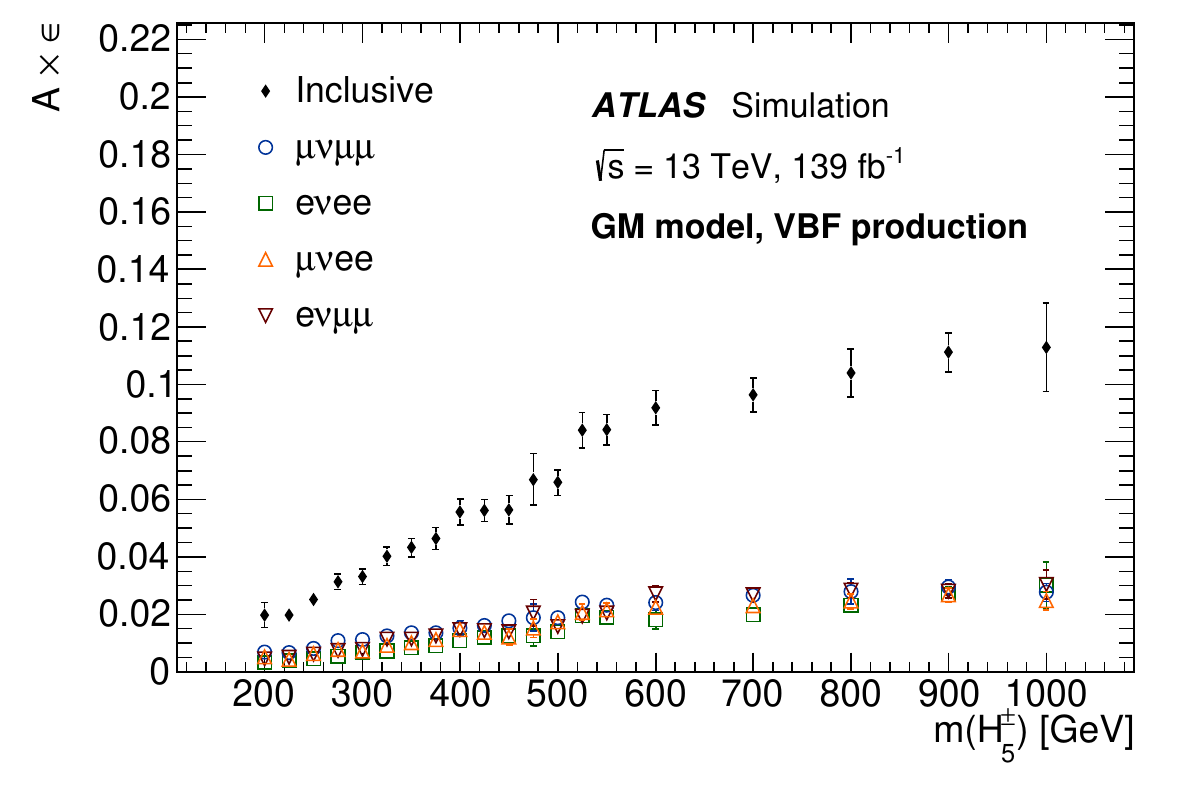}
\includegraphics[width=.7\textwidth]{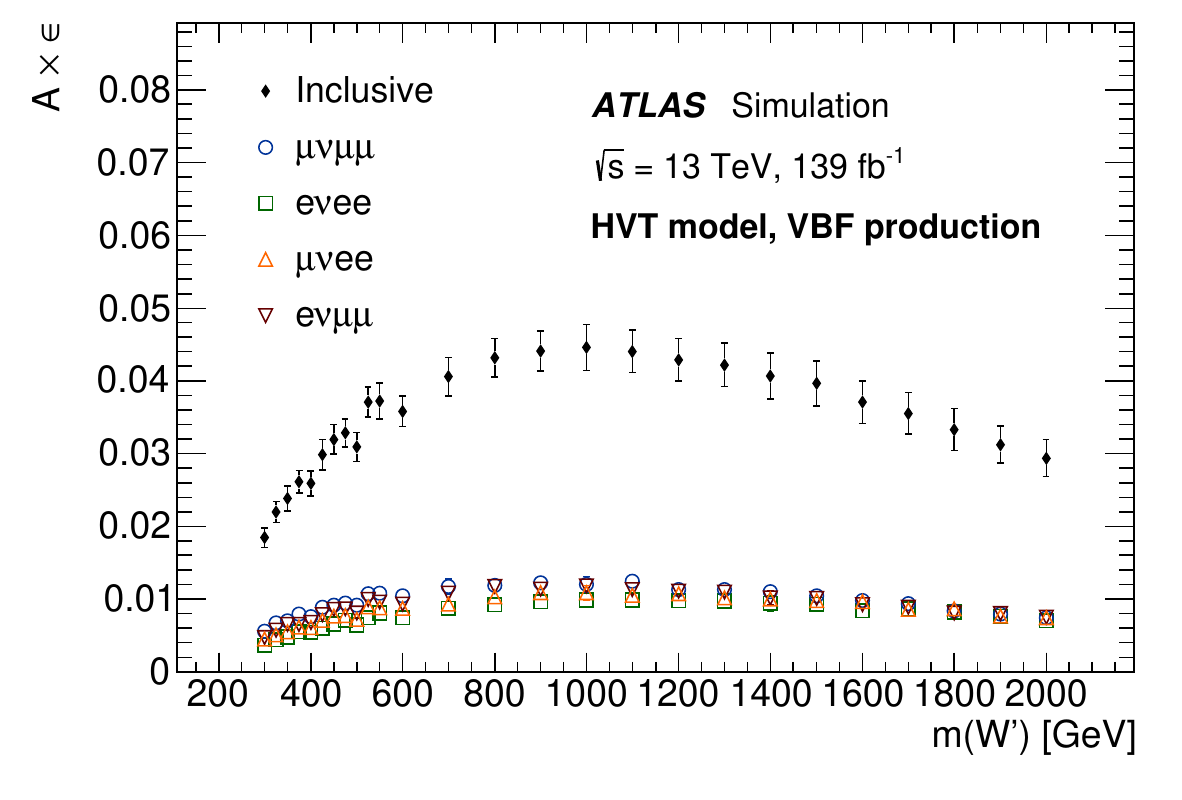}
\end{center}
\caption{The acceptance (A) times efficiency ($\epsilon$) of VBF  $H^{\pm}_{5}$ and HVT $W'$ selection after the ANN-based VBF selection at different mass points for the individual channels  $\mu\nu\mu\mu$,  $e\nu ee$, $\mu\nu ee$, $e\nu\mu\mu$, and the sum of all channels. The uncertainty includes both the statistical and experimental systematic components.}
\label{fig:Results:ANNGMAcceptance}
\end{figure}

The acceptance times efficiency $\cal{A\times\epsilon}$ of the ANN-based VBF selection as a function of the mass of the VBF $H^{\pm}_{5}$ and of the HVT $W'$ boson, relative to the generated signal events, is shown in Figure~\ref{fig:Results:ANNGMAcceptance}. There, decays of $W$ and $Z$ bosons into all flavours of charged leptons are included at event generation. For the $H^{\pm}_{5}$ and the HVT bosons the $\cal{A\times\epsilon}$ value is in the range  2–12\% and 2–5\% respectively for resonance masses ranging between 200 and 1000 GeV, the difference being due, with approximately equal importance, to the generator-level selection and the different angular distributions of the final products.

\section{Background estimation}

\label{sec:Background}
 
The backgrounds are classified into two groups: the irreducible backgrounds where all reconstructed lepton candidates are prompt (arise from the primary process) and the reducible backgrounds where at least one of the lepton candidates is not prompt. Non-prompt leptons are also referred to as "fake/non-prompt" leptons.
 
The contributions from the irreducible backgrounds \textit{WZ}-QCD, \textit{WZ}-EWK, \textit{ZZ}, $VVV$ and $t\bar{t}V$ are estimated using MC simulation. The normalizations of \textit{WZ}-QCD and \textit{ZZ} are constrained by data in a simultaneous fit using the signal region and dedicated control regions. Each signal region has two associated CRs designed to match that particular SR's event topology and jet multiplicity, as summarized in Table~\ref{tab:Selection}.
 
The dominant source of irreducible background is QCD-mediated production of \textit{WZ} dibosons.  Two CRs are created to constrain it. One is referred to as the \textit{WZ}-QCD Drell--Yan CR and is dedicated to the Drell--Yan analysis. It is defined by selecting the subsample of \textit{WZ} events that fulfill all the Drell--Yan event selection except the boson $\pt$ to resonance mass ratio. The lower bound on the boson $\pt$ to resonance mass ratio is set to $0.1$ to bring this CR close to the signal region. The second CR, referred to as the \textit{WZ}-QCD VBF CR, is dedicated to the VBF analyses. It is defined by selecting from the \textit{WZjj} subsample the events that fail the ANN output score requirement, and have $m_{\textit{jj}}>500$~GeV. The high $m_{\textit{jj}}$ requirement is applied in order to match the signal region's event topology. In both CRs the \textit{WZ}-QCD contribution is around 80\%.
 
To extract the \textit{ZZ} background normalization, two \textit{ZZ}-enriched control regions are defined after applying the \textit{WZ} event preselection described in Section~\ref{subsec:preselection}. The presence of at least a fourth lepton candidate satisfying the \textit{Baseline} identification criteria is required and no requirement on the missing transverse momentum is applied. This region is used for the Drell--Yan process analysis and is referred to as the \textit{ZZ} Drell--Yan CR. For the VBF selection, the events in the \textit{ZZ} VBF CR must have, in addition, at least two VBF tagged-jets.
 
The QCD-mediated production of \textit{ZZ} events represents 91\% (80\%) of the \textit{ZZ} Drell--Yan CR  (\textit{ZZ} VBF CR), while the contribution of \textit{ZZ} events  from electroweak-mediated production \textit{ZZ} is small. In the following the sum of the two components is referred as the  \textit{ZZ} background.
 
To validate the modelling of the  $t\bar{t}V$ background, a dedicated validation region is built by requiring the \textit{ WZjj} events to have at least one $b$-tagged jet. Since no significant data mis-modelling was observed, the $m$(\textit{WZ}) shape and normalization of this background are taken from simulation.

The reducible backgrounds originate from Drell--Yan $Z\to \ell \ell,~W\to\ell\nu$, $Z\gamma$, $t\bar{t}$, \textit{ Wt} and \textit{ WW} processes where jets or photons were misidentified as leptons. For both analysis regions the normalizations of the reducible backgrounds are estimated using a data-driven method. The method is based on a global matrix which exploits differences between the characteristics of real and fake/non-prompt leptons on a statistical basis. Details of the method can be found in Ref.~\cite{STDM-2014-02} . The shape in the Drell--Yan analysis is obtained from the data-driven method. In the VBF analysis, due to the fewer data events, the shapes are taken from simulation.


\section{Systematic uncertainties}

\label{Sec:Uncertainties}
 
Systematic uncertainties from the theoretical modelling and the object and event reconstruction have an impact in the signal and control regions used. The search sensitivity is then affected by their effects on background estimates, signal acceptance, and the shape of the distributions of the invariant mass discriminant. Depending on the nature of the uncertainty these can be classified into two groups:  (a) theoretical uncertainties associated with the MC modelling of both the background and signal processes and (b) experimental uncertainties related to the detector and reconstruction performance. The uncertainties and the methods used to evaluate them are discussed below. Unless explicitly stated, the uncertainties quoted are the uncertainties in the quantities themselves, not their impact on the search sensitivity.

\subsection{Theoretical uncertainties}
\label{subsec:TheoryUncert}
 
Systematic uncertainties in the theoretical modelling by the event generators used to evaluate the \textit{WZ}-QCD, \textit{WZ}-EWK and \textit{ZZ} templates are considered. For the \textit{WZ}-QCD and \textit{ZZ} backgrounds that have data-driven normalization only the shape variations of the reconstructed $m$(\textit{WZ}) distribution are considered. Uncertainties due to higher-order QCD corrections are evaluated by varying the renormalization and factorization scales independently by factors of two and one-half. For the \textit{WZ}-QCD background only a small shape effect is observed and used in the fit. For the \textit{WZ}-EWK background the uncertainties in the $m$(\textit{WZ}) shape grow with the mass from 8\% to 15\%. The uncertainties due to the PDF and the $\alpha_s$ value used in the PDF determination are evaluated using the PDF4LHC prescription~\cite{Butterworth:2015oua}. For the \textit{WZ}-QCD background they are estimated to have a small shape component but are nevertheless included in the fit. For the \textit{WZ}-EWK they are added in quadrature, and the total uncertainty stays between 5 to 6\% in all mass bins for both the Drell-Yan and VBF selections. A modelling uncertainty in the \textit{WZ}-QCD background template including effects of the parton shower model, is estimated by comparing predictions of the $m$(\textit{WZ}) distribution from the \textsc{Sherpa} and \textsc{MadGraph} MC generators.  The difference between the two predicted $m$(\textit{WZ}) distribution shapes is used as an uncertainty band centred around the nominal \textsc{Sherpa} prediction.
A parton shower modelling uncertainty in the \textit{WZ}-EWK background template is estimated using two MC samples with different parton shower models, \textsc{Pythia} and \textsc{Herwig}. This modelling uncertainty has no effect on the $m$(\textit{WZ}) distribution's normalization at low mass, but grows to 5\% at high mass.
 
For the \textit{ZZ} background the shape uncertainties originating from the renormalization and factorization scales, as well as from the PDF and the chosen value of $\alpha_s$ are evaluated in a similar way.
 
The theoretical uncertainties described above are evaluated in all the analysis signal and control regions and treated as uncorrelated across those regions in the statistical analysis.
 
An uncertainty of 20\% is assigned to the $t\bar{t}V$ and $VVV$ cross-sections \cite{ATL-PHYS-PUB-2016-002,TOPQ-2016-11,STDM-2017-22}. It consists of contributions from PDF uncertainties and QCD scale uncertainties.

Uncertainties in the signal acceptances due to PDF and scale choices are also evaluated. These uncertainties are calculated following the procedure described above, for several resonance mass points, and for each model, production process and decay. The theoretical uncertainties of the HVT signals are evaluated to be less than 20\% for all production modes and they are 30\% for the GM model.

\subsection{Experimental uncertainties}
\label{subsec:ExpUncert}
 
Experimental uncertainties arise from the determination of the luminosity, the lepton trigger efficiency, the reconstruction and identification efficiencies of leptons and jets, and the missing transverse momentum.

The uncertainty in the integrated luminosity is 1.7\%. It is derived following a methodology similar to that detailed in Ref.~\cite{ATLAS-CONF-2019-021}, using the LUCID-2 detector for the baseline luminosity measurements~\cite{Avoni_2018}, and calibrating the luminosity scale using $x$-$y$ beam-separation scans. A variation in the \pileup reweighting of MC events is included to cover the uncertainty in the ratio of the predicted and measured inelastic cross-sections~\cite{STDM-2015-05}.

Systematic uncertainties affecting the reconstruction and energy calibration of jets are propagated through the analysis. They are the dominant experimental uncertainties in the VBF selection.
Those due to the jet energy scale and resolution are obtained from simulations and in situ techniques~\cite{PERF-2016-04}. The uncertainties in the $b$-tagging efficiency and the mis-tag rate are also taken into account.  The effect of jet uncertainties on the expected number of events ranges up to 15\% in the VBF selection.
 
Uncertainties in the efficiencies of the lepton triggers are found to be negligible. The uncertainties due to the electron and muon reconstruction, identification and isolation requirements are estimated using tag-and-probe methods in $Z\rightarrow \ell \ell$ events in data and simulation~\cite{EGAM-2018-01,PERF-2015-10}. Uncertainties in the lepton energy scale and resolution are also assessed. The impact of lepton uncertainties on the expected number of events is typically below 1\%.
 
The uncertainty in the measurement of missing transverse momentum is estimated by propagating the uncertainties in the transverse momenta of preselected leptons and jets, as well as those in the soft term~\cite{PERF-2016-07}.

An uncertainty in the prediction of the fake/non-prompt background is also taken into account because it affects the shape and normalization of the background distributions. The total uncertainty is about 60\% (more than 100\%) for the Drell--Yan (VBF) selections. It is larger for the VBF selection because of the higher statistical uncertainty.


\section{Results}

\label{sec:Results}
 
\subsection{Statistical analysis strategy}

The \textit{WZ} invariant mass distribution is used as the discriminating variable. The bin widths were chosen to be comparable to the expected resolution for the resonance model under investigation, and at the same time to optimize the sensitivity of the search while reducing the impact of statistical fluctuations.
 
A profile-likelihood-ratio test statistic~\cite{Cowan:2010js} is used to test the compatibility of the background-only hypothesis with the data and to test the signal-plus-background hypothesis. The binned likelihood function is constructed by considering, in each bin, the contributions of the backgrounds and of a hypothetical signal of given strength relative to a benchmark model's production cross-section. In the absence of an observed signal, exclusion limits on the presence of a signal are then derived using the CLs method~\cite{Read:2002hq}. All lepton flavours and data taking periods are combined together for the profile-likelihood-ratio test since there was no significant gain in splitting the samples before the statistical analysis.
 
Simultaneous maximum-likelihood fits to the observed binned distributions of $m$(\textit{WZ}) in the signal regions and their dedicated \textit{WZ} and \textit{ZZ} control regions are performed. Separate fits are performed for the Drell--Yan and VBF selections.
The normalizations of the \textit{WZ}-QCD and \textit{ZZ} contributions are freely floating parameters in these fits and are constrained by the data in both the SRs and dedicated CRs. The ratio of the fitted contributions in the CR and SR is allowed to vary within the theoretical uncertainties. The normalizations and shapes of all other backgrounds are allowed to vary within their uncertainties.
 
Systematic uncertainties, described in Section~\ref{Sec:Uncertainties}, and their correlations are incorporated in the likelihood as nuisance parameters with
Gaussian constraints. Most of the systematic uncertainties are taken to be correlated between the SR and CRs and fit simultaneously in these regions, with the theoretical uncertainties of the \textit{ZZ}, \textit{WZ}-QCD and \textit{WZ}-EWK backgrounds being the only exceptions.
 
Two fit configurations are used, referred to as the Drell--Yan and VBF configurations. The Drell--Yan fits include the Drell--Yan SR, \textit{WZ}-QCD CR and \textit{ZZ}-CR. In the VBF configuration, fits include the VBF-SR, \textit{WZjj}-QCD CR and \textit{ZZjj}-CR. Separate fits are performed for the different models tested and for different resonance mass hypotheses. The Drell-Yan configuration is used to search for a $W'$ boson predicted by the HVT benchmark. Two VBF fits are performed using the VBF configuration: one for the search for a VBF-produced $W'$ predicted by the HVT model, and the other for the search for a charged Higgs boson, $H_5^{\pm}$, as predicted by the GM model.
 
\subsection{Data and background comparisons}
 
To test the compatibility of the data and the background expectations,
the data are first fit to the background-only hypothesis, separately in the Drell--Yan and VBF configurations.
 
The post-fit background yields are summarized in Table~\ref{table::Post-fitResultsSignal} for the Drell--Yan and VBF signal regions. In both cases the fit is able to adjust the SM \textit{ZZ} and \textit{WZ}-QCD background normalizations using the data in signal and control regions. In the Drell--Yan fit, the \textit{ZZ} background normalization is increased by around 10\% while the \textit{WZ}-QCD background is decreased by 10\% relative to the pre-fit predictions. Some mild pulls in the modelling uncertainties by less than one standard deviation from their pre-fit values are visible in the Drell--Yan fit. In the VBF fits, the normalization of the \textit{ZZ} background is consistent with the pre-fit value while the \textit{WZ}-QCD background is reduced by around 30\%. Apart from the mild pulls in the Drell--Yan signal region, none of the other nuisance parameters are significantly pulled or constrained relative to their pre-fit values in any of the background-only fits.
 
\begin{table}[h!]
\caption{Expected and observed yields in the Drell--Yan and VBF signal regions. The yields and uncertainties are presented
after the background-only fit to the data in the Drell--Yan or VBF signal regions. The
uncertainty in the total background estimate is smaller than the sum in quadrature of the individual background
contributions due to anti-correlations between the estimated backgrounds from different sources.}
\label{table::Post-fitResultsSignal}
\centering
\begin{tabular}{l|c|c}
\hline
\hline
& Drell-Yan signal region &  VBF signal region \\
\hline
\textit{WZ}-QCD          & $1734 \pm 77$ & $29 \pm 4$ \\
\textit{WZ}-EWK          & $89   \pm 10$  & $26 \pm 3$ \\
$VVV + t\bar{t}V$     & $148  \pm 27$ & $0.9 \pm 0.2$ \\
\textit{ZZ}              & $95   \pm 5$  & $5 \pm 1$ \\
Fakes/non-prompt leptons  & $88   \pm 49$ & $0.3 \pm 0.8$\\
\hline
Total background & $2155 \pm 71$  & $61 \pm 6$ \\
Observed & 2155 & 66 \\
\hline
\hline
\end{tabular}
\end{table}

\begin{figure}[htbp]
\centering
\subfloat[]{
\includegraphics[width= 0.45\textwidth]{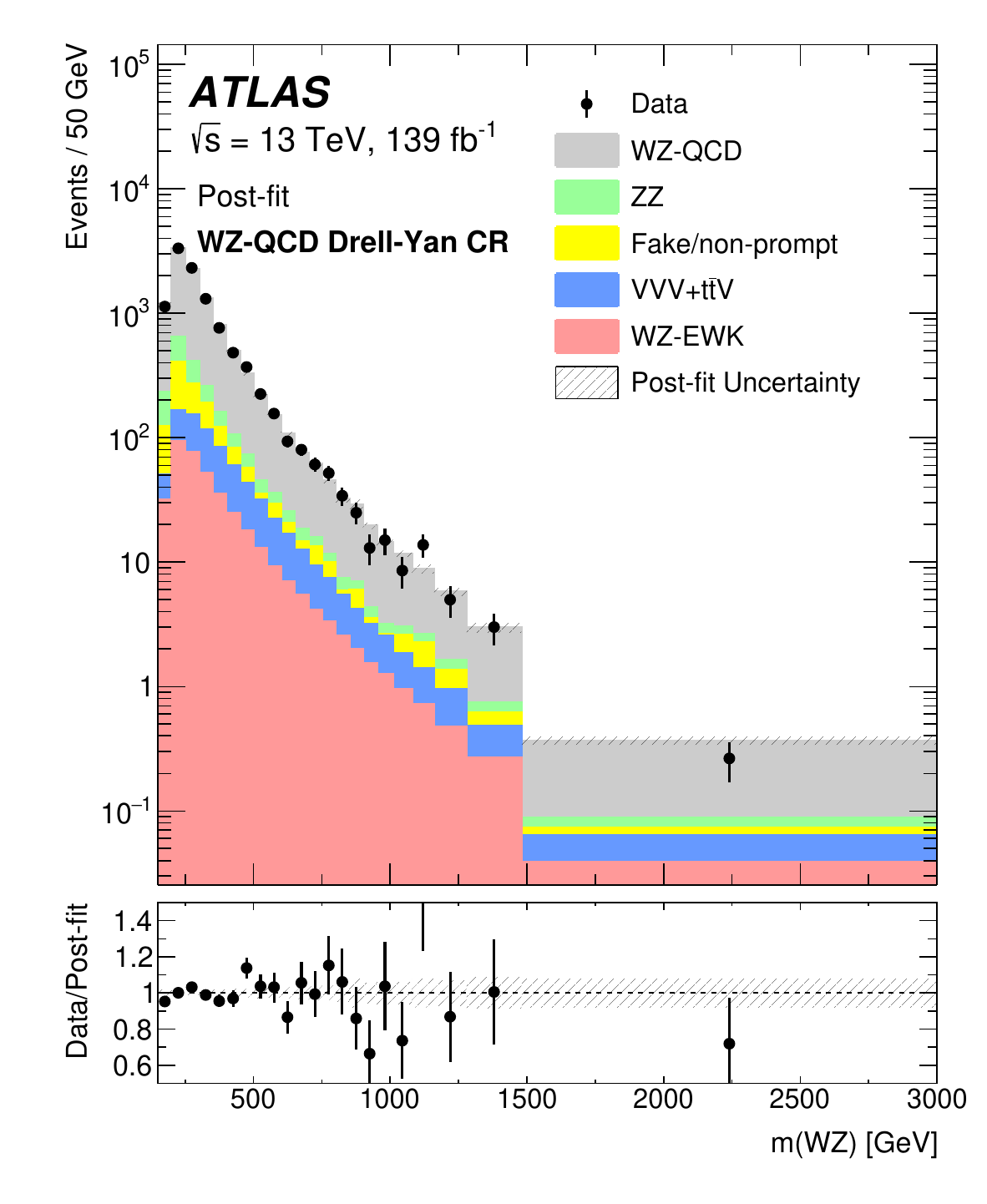}
\label{fig:Results:Postfitqq:WZCR}
}
\subfloat[]{
\includegraphics[width= 0.45\textwidth]{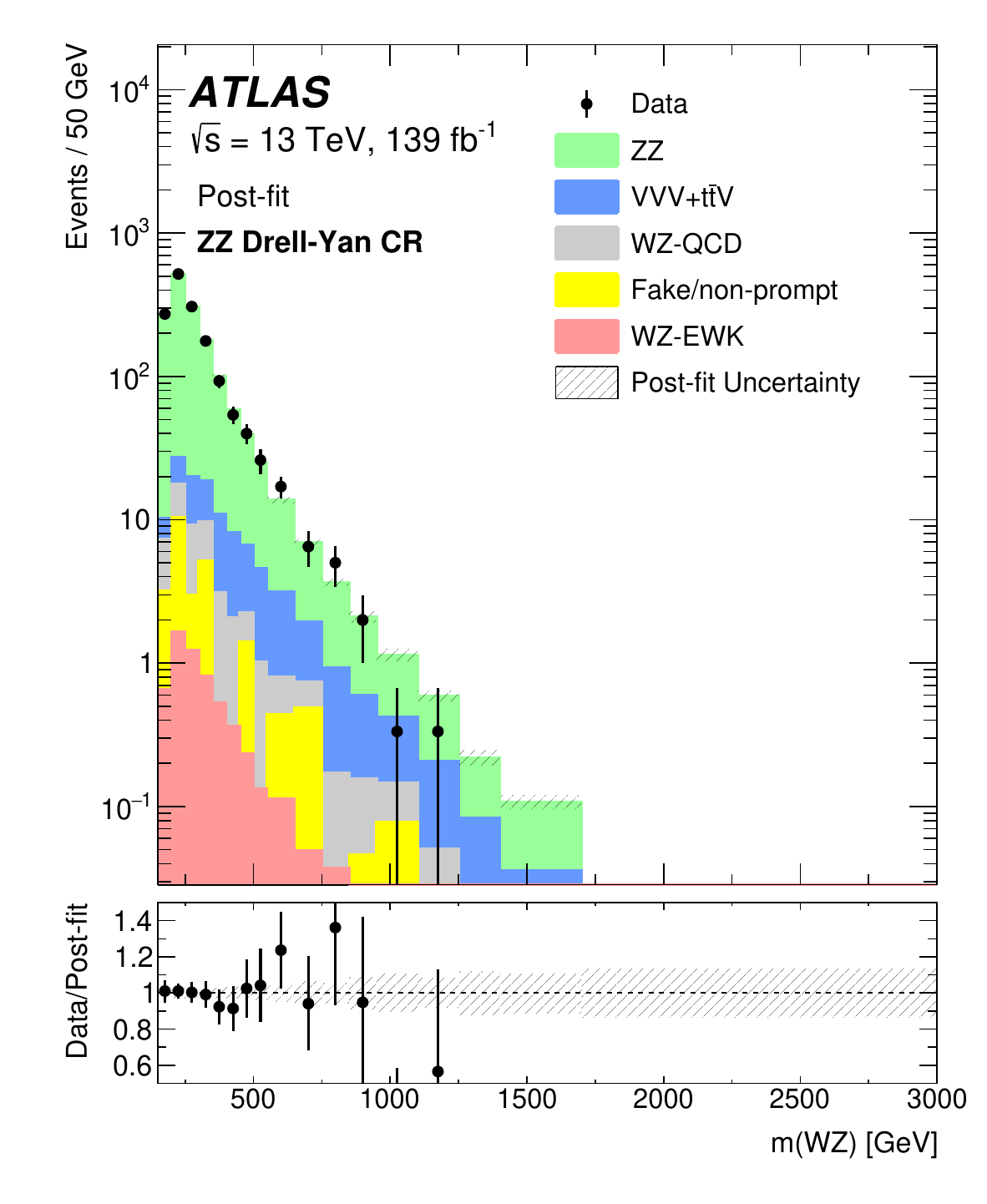}
\label{fig:Results:Postfitqq:ZZCR}
}\\
\subfloat[]{
\includegraphics[width=0.45\textwidth]{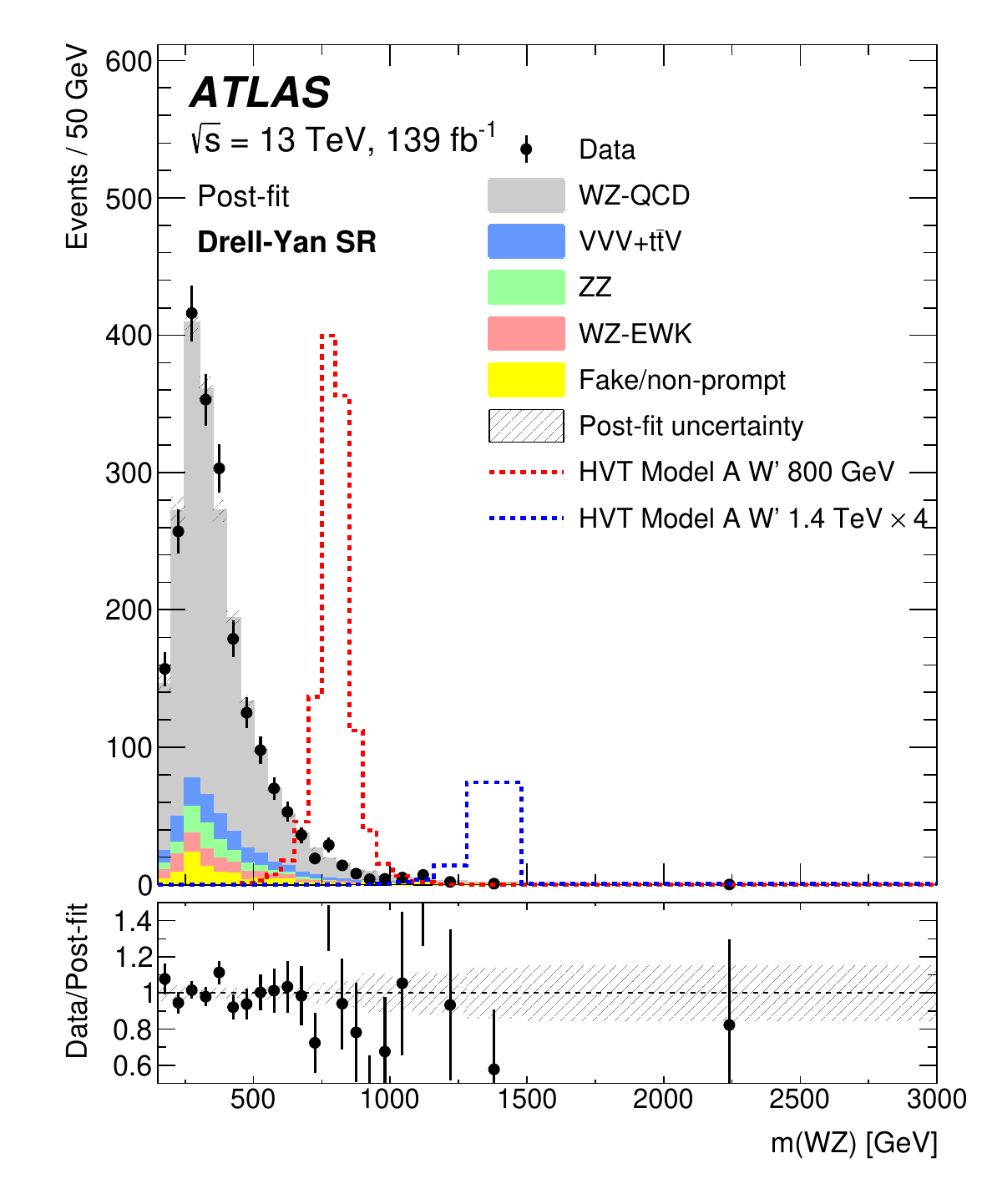}
\label{fig:Results:Postfitqq:SR}
}
\caption{Comparisons of the data and the expected background distributions of the \textit{WZ} invariant mass in the (a) \textit{WZ}-QCD,  (b) the \textit{ZZ} control regions and (c) the Drell--Yan signal region. The background predictions are obtained from a background-only simultaneous fit to the Drell-Yan signal region and its control regions. For illustration, the expected distributions from HVT $W'$ resonances (model A) with masses of 800~\GeV~and 1.4~\TeV, with the latter normalized to four times its predicted cross-section, are shown in the signal region.
The bottom panels show the ratios of the data to the post-fit background predictions. The uncertainty in the total background prediction, shown as  a hashed area, combines statistical and systematic contributions. }
\label{fig:Results:qqPostfit}
\end{figure}

\begin{figure}[htbp]
\centering
\subfloat[]{
\includegraphics[width= 0.45\textwidth]{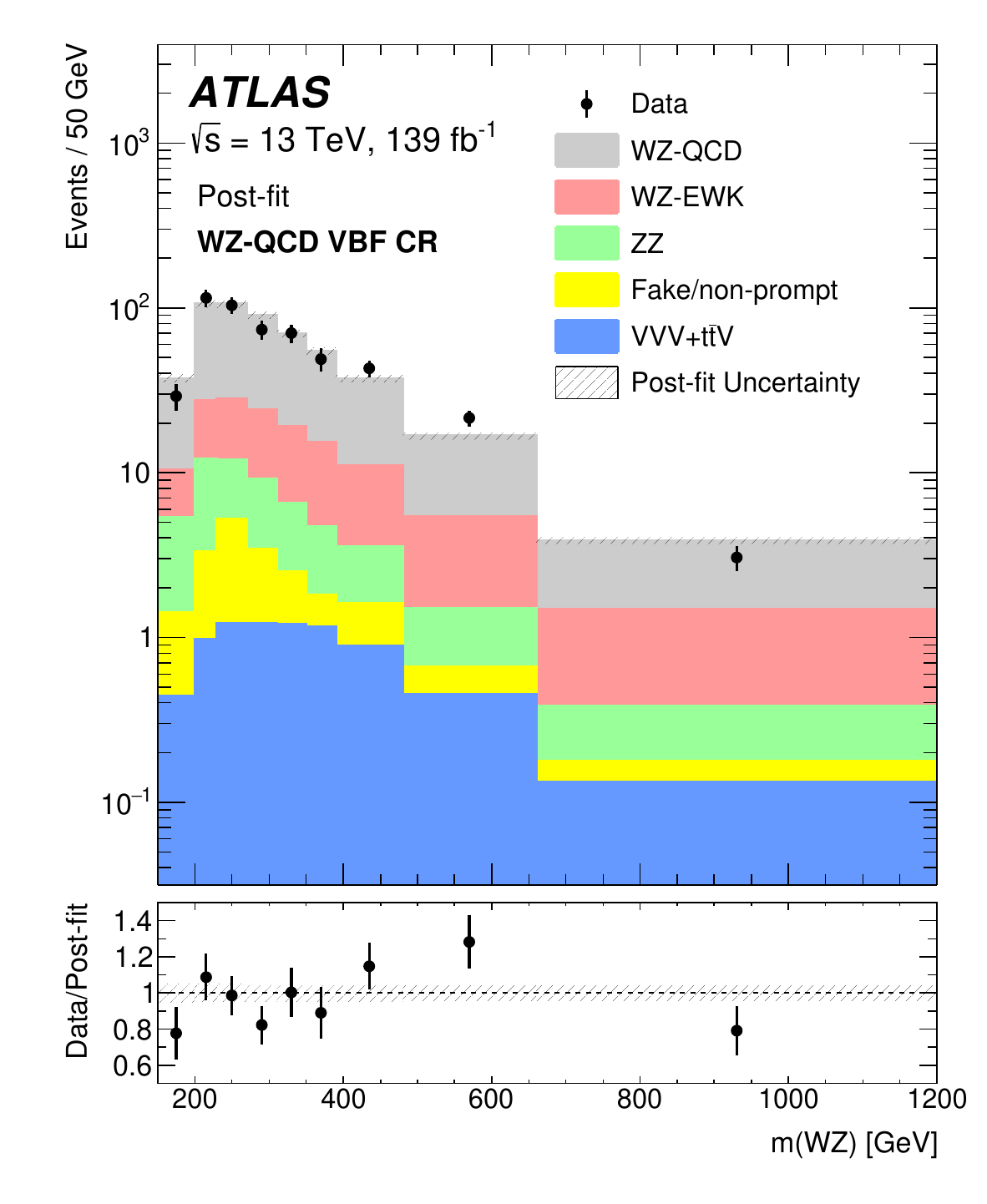}
\label{fig:Results:Postfitvbf:WZCR}
}
\subfloat[]{
\includegraphics[width= 0.45\textwidth]{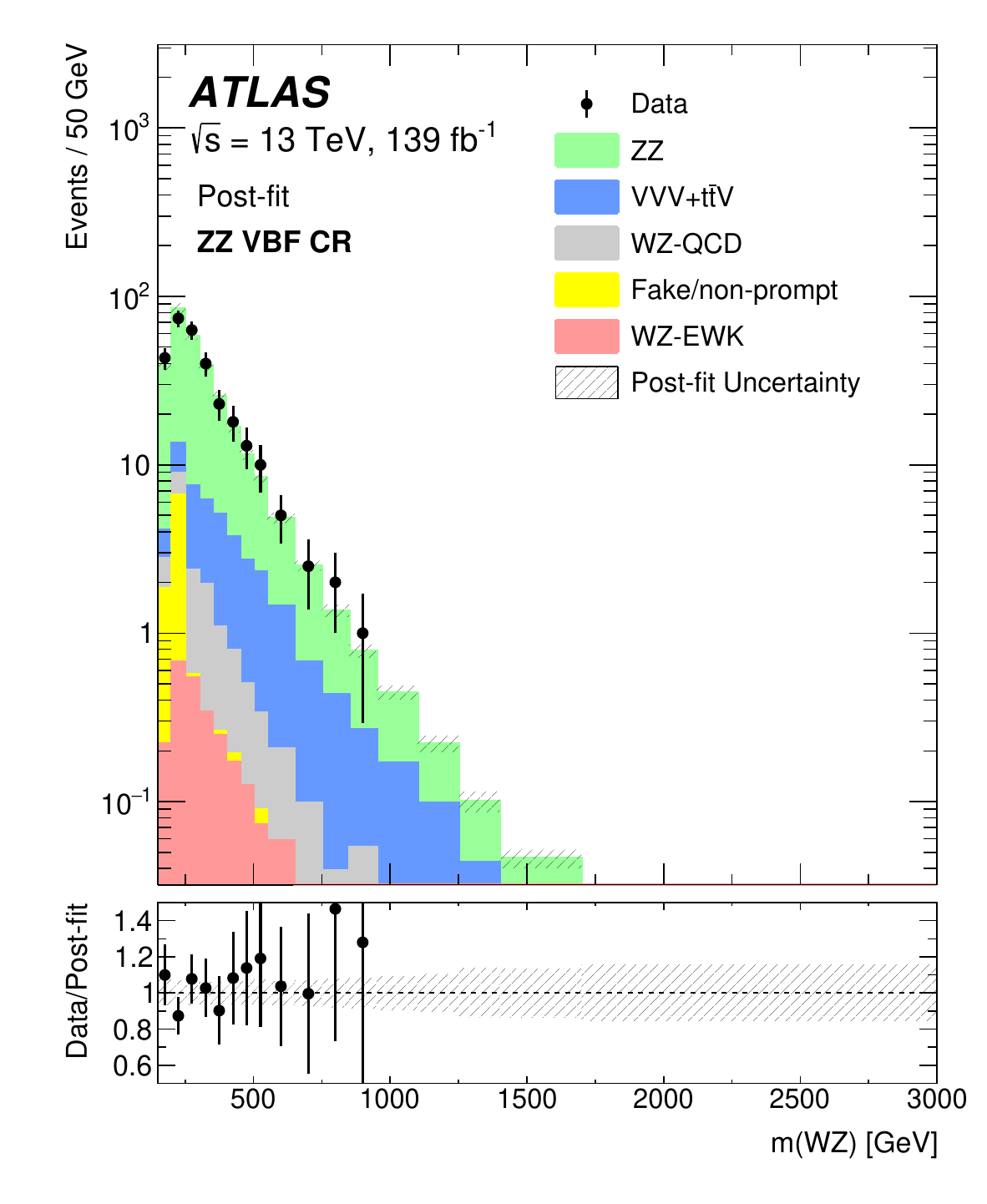}
\label{fig:Results:Postfitvbf:ZZCRJJ}
}\\
\subfloat[]{
\includegraphics[width=0.45\textwidth]{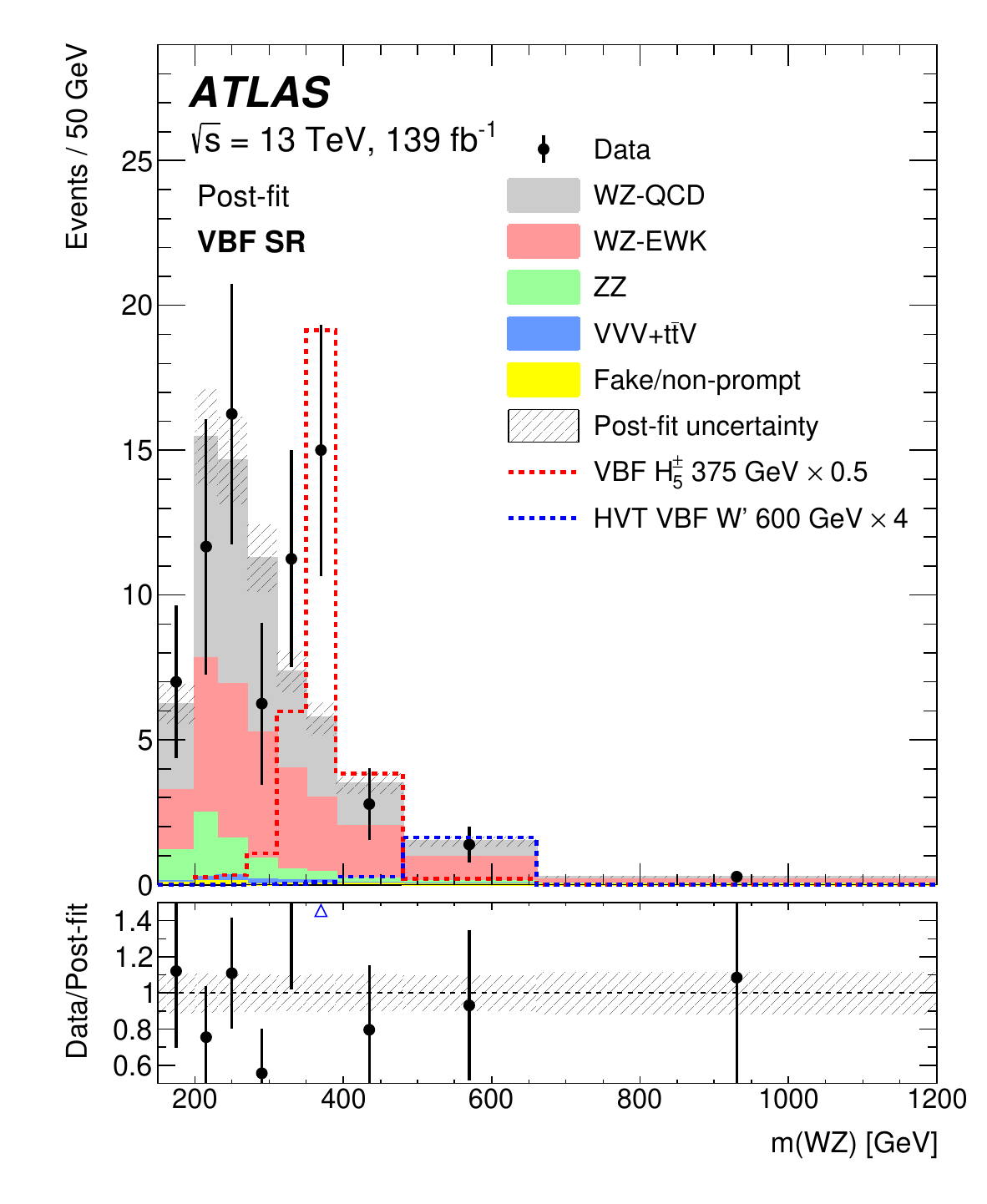}
\label{fig:Results:Postfitvbf:SR}
}
\caption{Comparisons of the data and the expected background distributions of the \textit{WZ} invariant mass in
(a) the \textit{WZ}-QCD VBF control region, (b) the \textit{ZZ} VBF control region and (c) the VBF signal region. The background predictions are obtained from a background-only simultaneous fit to the VBF signal region and its two control regions.
For illustration, the expected distributions from an $H_5^{\pm}$ GM model resonance ($\sin\theta_H= 0.5$) with a mass of 375~\GeV~and from an HVT $W'$ (model A) of mass 600~\GeV~are shown in the signal region, with the predicted cross sections scaled by 0.5 and 4, respectively. The bottom panels show the ratios of the data to the post-fit background predictions. The uncertainty in the total background prediction, shown as a hashed area, combines statistical and systematic contributions. }
\label{fig:Results:GMPostfit}
\end{figure}

The post-fit $m$(\textit{WZ}) distributions in the signal regions and their respective \textit{WZ}-QCD and \textit{ZZ} control regions are shown in Figure~\ref{fig:Results:qqPostfit} for the Drell--Yan selection and in Figure~\ref{fig:Results:GMPostfit} for the VBF selection.
The bottom panels show that the observed mass distributions are in good agreement with the estimated post-fit background contributions in all signal and control regions.
 
The largest observed excess is in the VBF category at $m$(\textit{WZ}) around 375~GeV, as shown in Figure~\ref{fig:Results:Postfitvbf:SR}.
The local significances for VBF produced signals of a charged Higgs boson $H_5^{\pm}$ or an HVT $W'$ boson are 2.8 and 2.5 standard deviations, respectively. The respective global significances calculated using the look-elsewhere effect as in Ref.~\cite{ATL-PHYS-PUB-2011-011}, and evaluated up to a mass of 1.2~TeV, are 1.6 and 1.7 standard deviations. In the Drell--Yan signal region the largest difference between the data and the SM background prediction is located around a mass of 1.1~TeV with a local significance of 1.2 standard deviations.

\subsection{Impact of systematic uncertainties}
 
The effects of systematic uncertainties on the search are studied for hypothesized signals using a signal-strength parameter $\mu$, which is the ratio of the extracted cross-section to the injected hypothesized signal cross-section.
For this study, the signal production cross-section is set to be equal to the expected median upper limits (Section~\ref{subsec:Limits}). The expected relative uncertainties in the best-fit $\mu$ value after the maximum-likelihood fit are shown in Table~\ref{table:Impact} for two reference models and mass points: Drell--Yan production of a $W'$ boson in the HVT model with mass $m(W')$~=~1100~GeV, and VBF production of an $H^{\pm}_5$ in the GM model with mass $m(H^{\pm}_5)$~=~375~GeV.
The individual sources of systematic uncertainty are combined into fewer background modelling and experimental categories. For signals with higher mass, the data statistical uncertainty is dominant. The uncertainties with the largest impact on the sensitivity of the searches are from the normalization of the irreducible backgrounds \textit{WZ}-QCD and \textit{ZZ},  from the theory modelling of the \textit{WZ} background (in the table this includes  QCD and EWK components), from the reducible background shape and normalization, and from the sizes of the MC samples. Uncertainties related to luminosity and \pileup also play a relevant role in both signal regions. The jet uncertainties, such as those in the jet energy scale and resolution, naturally have a large impact in the VBF search.
 
\begin{table}[h!]
\caption{
Dominant relative uncertainties in the best-fit signal-strength parameter ($\mu$) for a hypothetical HVT signal of mass $m(W')$~=~1\,100~GeV in the Drell-Yan signal region and a GM signal of mass $m(H^{\pm}_5)$~=~375~GeV in the VBF signal region.
For this study, the production cross-sections of the signals are set to the expected median upper limits at these two mass values. Uncertainties with smaller contributions are not included.}
\centering
\begin{tabular}{l|c|c}
\hline
\hline
Source of uncertainty & \multicolumn{2}{c}{$\Delta\mu/\mu$ [$\%$]}\\
\hline
& Drell--Yan signal region & VBF signal region\\
& $m(W')$~= 1100 GeV          & $m(H^{\pm}_5)$~= 375 GeV\\
\hline
\textit{WZ}-QCD + \textit{ZZ} normalization  & 2 & 11\\
\textit{WZ} background: parton shower & 6 & 1\\
\textit{WZ} background: scale, PDF    & 5 & 8\\
Fake/non-prompt background     & 3 & 1\\
\textit{ZZ} background: scale, PDF      & 0.2 & <0.1 \\
$VVV+t\bar{t}V$ modelling      & 3 & 1 \\
\hline
Electron identification                   & 6 & 3\\
Muon identification                       & 1 & 4\\
Jet uncertainty                           & 0.8 & 16\\
Flavour tagging                           & 0 & 1\\
Missing transverse momentum               & 0.2 & 0.5\\
MC statistical uncertainty                & 10 & 5\\
Luminosity                                & 2 & 8\\
Pile-up                                    & 0.1 & 8\\
\hline
Total systematic uncertainty             & 16 & 22\\
Data statistical uncertainty            & 54 & 55\\
Total                                    & 56 & 59\\
\hline
\hline
\end{tabular}
\label{table:Impact}
\end{table}

\subsection{Limits on the production of heavy resonances }
\label{subsec:Limits}
 
Constraints on the production of heavy resonances are derived by repeating the test of the signal-plus-background hypothesis for different signal models. Upper limits on cross-sections times branching fraction to \textit{WZ} are calculated using the asymptotic approximation~\cite{Cowan:2010js}.
 
\begin{figure}[!htbp]
\begin{center}
\includegraphics[width=.7\textwidth]{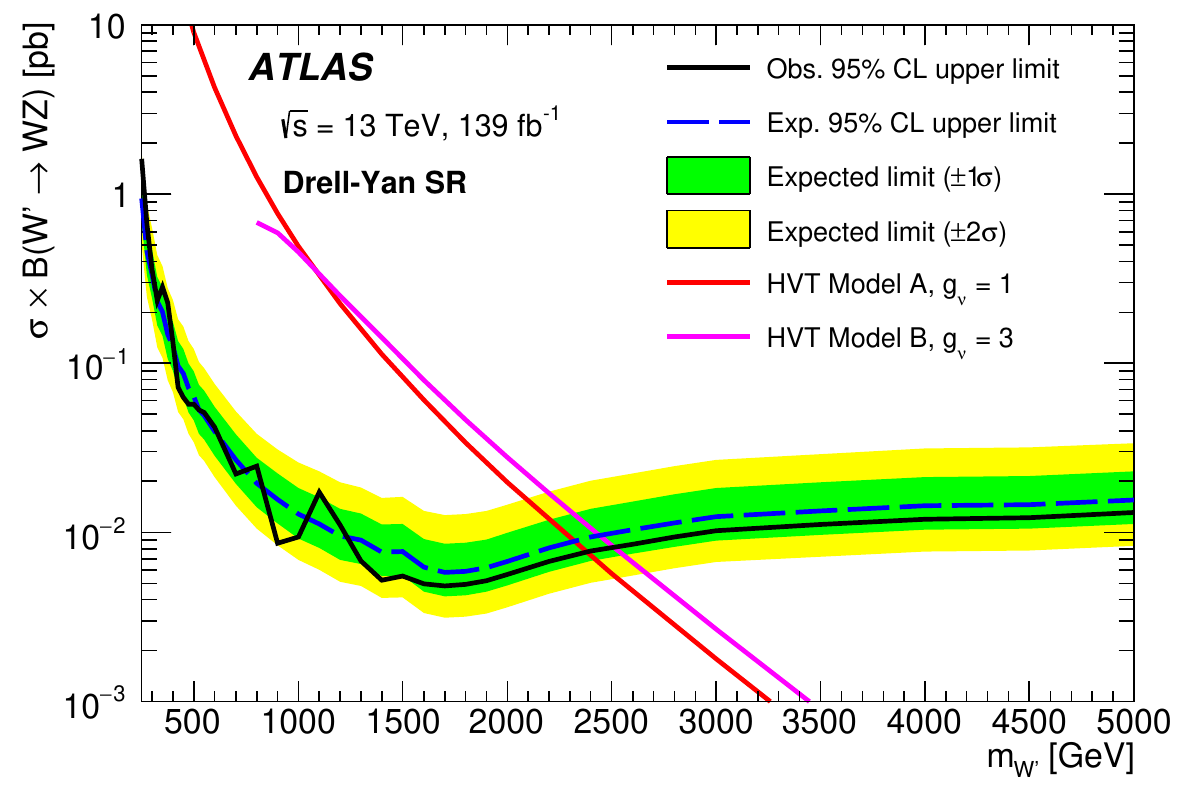}
\end{center}
\caption{Observed and expected 95\% CL exclusion upper limits on $\sigma \times B(W' \rightarrow{WZ})$ for the Drell--Yan production of a $W'$ boson in the HVT model as a function of its mass. The LO theory predictions for HVT Model A with $g_V=1$ and Model B with $g_V=3$ are also shown.}
\label{fig:Results:qqLimits}
\end{figure}

\begin{figure}[!htbp]
\begin{center}
\includegraphics[width=.7\textwidth]{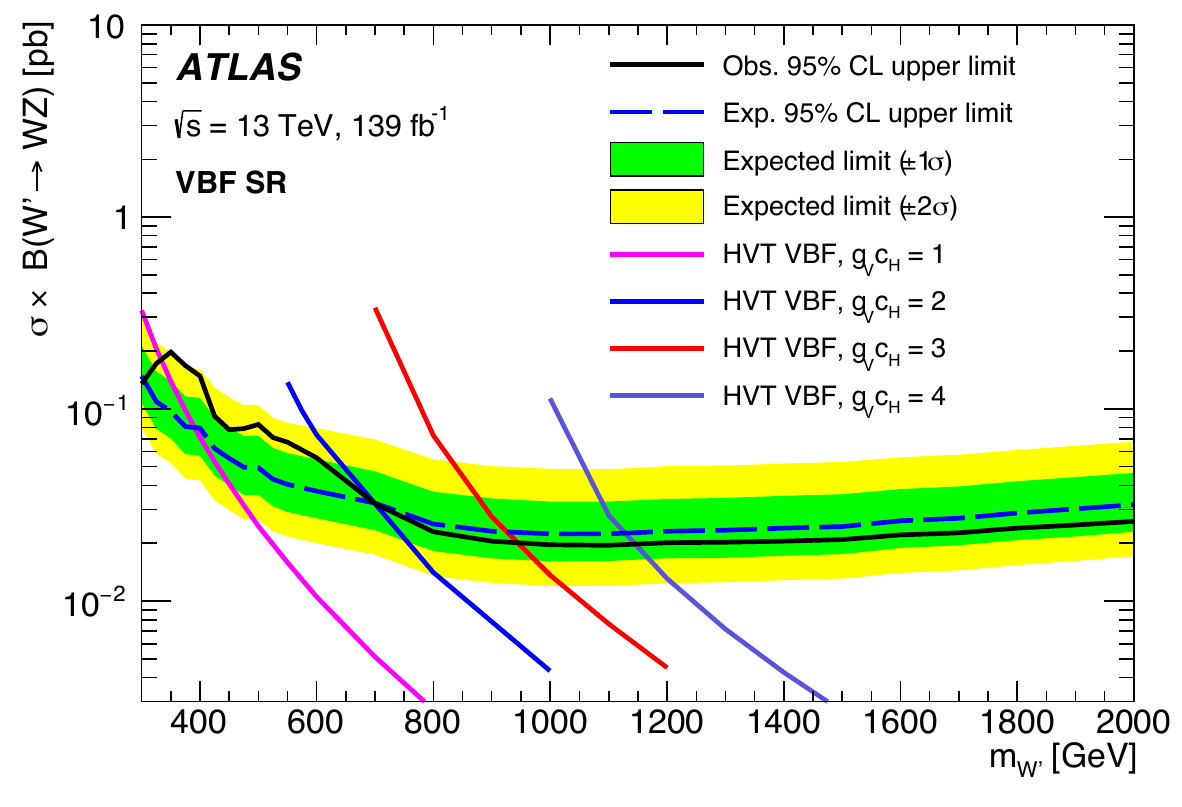}
\end{center}
\caption{Observed and expected 95\% CL upper limits  on $\sigma \times B(W' \rightarrow{WZ})$ for the VBF production of a $W'$ boson in the HVT benchmark with parameter $c_F$~=~0, as a function of its mass. The LO theory predictions for HVT VBF benchmark with different values of the coupling parameters $g_V$ and $c_H$ are also shown.}
\label{fig:Results:HVTLimits}
\end{figure}

For the HVT model search, Figure~\ref{fig:Results:qqLimits} presents the observed and expected limits on $\sigma \times B(W' \rightarrow{WZ})$ at 95\%~CL as a function of the $W'$ mass for the HVT model in the Drell--Yan signal region. Masses below 2.4~TeV can be excluded for Model A and 2.5~TeV for Model B. For resonance masses above 2~TeV the exclusion limits become weaker due to the poorer acceptance  at high mass (see Figure~\ref{fig:Results:DrellYannAcceptance}).
Regarding the VBF production mode, 95\% CL limits on $\sigma \times B(W' \rightarrow{WZ})$ are shown in Figure~\ref{fig:Results:HVTLimits} for the benchmark model with $c_F=0$ and different values of $g_Vc_H$.
Thus, they close a gap left by the limits obtained by searches in the Drell-Yan process as $c_F$ approaches zero.
 
Because of large mass mixing, which depends on the coupling $g_Vc_H$,  between the SM gauge bosons and $V'$, the theory curves start at different values of V' mass, close to the limit of validity of the HVT model.
Masses below 340~\GeV, 700~GeV, 945~GeV and 1145~GeV can be excluded for the HVT VBF model with $c_F = 0$ and $g_V c_H = 1.0$, 2.0, 3.0 and 4.0, respectively.

\begin{figure}[htbp]
\centering
\subfloat[]{
\includegraphics[width=0.7\textwidth]{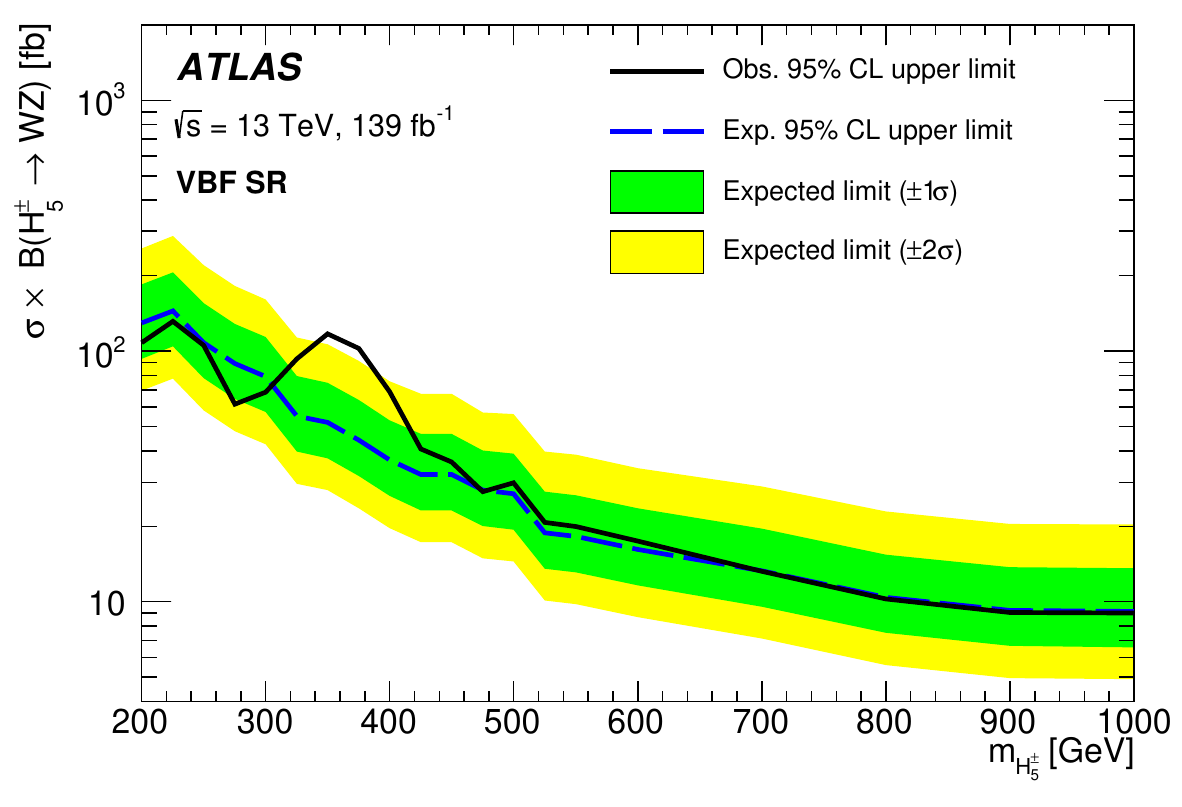}
\label{fig:Results:GMLimits:a}
}\\
\subfloat[]{
\includegraphics[width= 0.7\textwidth]{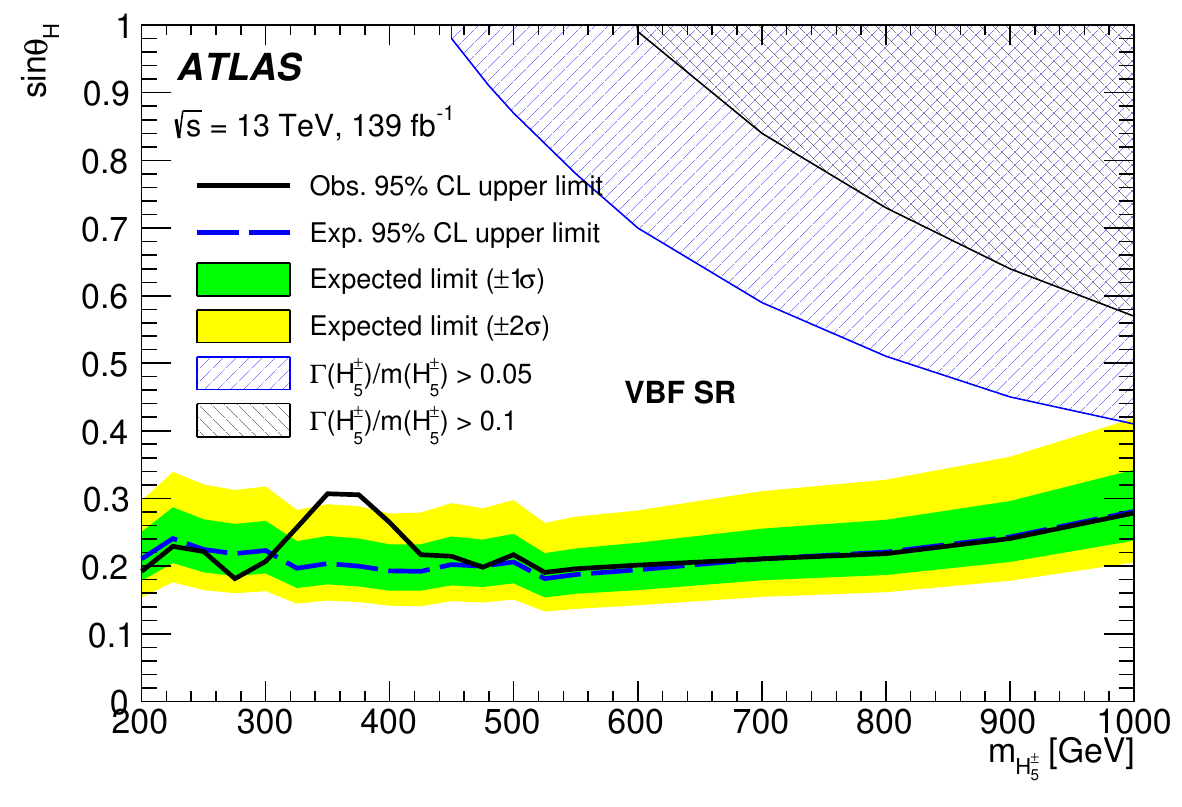}
\label{fig:Results:GMLimits:b}
}
\caption{Observed and expected 95\% CL upper limits \protect\subref{fig:Results:GMLimits:a} on $\sigma \times B(H^{\pm}_5 \rightarrow{WZ})$  and \protect\subref{fig:Results:GMLimits:b} on the parameter $\mathrm{sin} \theta_H$  of the GM model as a function of $m_{H^{\pm}_5}$. The shaded region shows where the theoretical intrinsic width of the resonance would be larger than 5\% or 10\% of the mass.}
\label{fig:Results:GMLimits}
\end{figure}

For the $H^{\pm}_5$ GM search, observed and expected exclusion limits at 95\%  CL on $\sigma \times B(H^{\pm}_5 \rightarrow{WZ})$  and on the mixing parameter $\sin\theta_H$ are shown in Figure~\ref{fig:Results:GMLimits}.
The latter  are about 35\% stronger than in the previous publication~\cite{EXOT-2016-11}. They are comparable to the CMS results~\cite{CMS-HIG-20-017}, based on the combination of searches for singly and doubly charged members of the $H_5$ fiveplet.
The shaded regions show the parameter space for which the $H^{\pm}_5$  width exceeds 5\% and 10\% of $m_{H^{\pm}_5}$.
The intrinsic width of the scalar resonance is narrower than the detector resolution in the mass region explored.

As a test of the asymptotic approximation used in the statistical
analysis, limits are also computed with ensembles of pseudo-experiments in all signal and control regions. The cross-section upper limits
obtained in that case agree in all cases for masses below 500~\GeV. At higher masses, where event yields become smaller, the discrepancy between the two methods become larger, but they remain within 6\%-10\%.
 
\FloatBarrier


\section{Conclusions}

\label{sec:Conclusions}
 
A search was performed for resonant $WZ$  production in fully leptonic final states  (electrons and muons) using 139~fb$^{-1}$ of $\sqrt{s} = 13~\TeV$ $pp$ collision data collected by the ATLAS experiment at the LHC.
Two different production processes are considered, Drell-Yan and vector-boson fusion.
 
The data in the Drell-Yan selection are found to be consistent with Standard Model predictions.  The results are used to derive upper limits at 95\% confidence level on the cross-section times branching ratio
for heavy vector triplet production in benchmark Model A (Model B) with coupling constant $g_V=1$ ($g_V=3$)
as a function of the resonance mass, with no evidence of heavy vector resonance production
for masses below 2.4 (2.5)~\TeV.
 
In the case of the VBF production process, limits on the production cross-section times branching ratio to $WZ$ of a hypothetical resonance  are obtained as a function of the mass for a  heavy vector triplet or for a charged member of the fiveplet scalar in the Georgi--Machacek model.
Values of the parameter $\sin \theta_H > 0.3$ are excluded for masses between 200 GeV and 1 TeV.
The results show a local excess of events over the Standard Model expectations at a resonance mass of around 375 GeV. The local significances for signals of a heavy vector $W'$ boson or a $H_5^{\pm}$ are 2.5 and 2.8 standard deviations respectively. The respective global significances
calculated considering the look-elsewhere effect are 1.7 and 1.6 standard deviations respectively.
With no evidence of heavy $W'$ vector-resonance production, limits on the production times branching ratio for the heavy vector triplet VBF production process have been obtained as a function of mass.


\section*{Acknowledgments}


We thank CERN for the very successful operation of the LHC, as well as the
support staff from our institutions without whom ATLAS could not be
operated efficiently.
 
We acknowledge the support of
ANPCyT, Argentina;
YerPhI, Armenia;
ARC, Australia;
BMWFW and FWF, Austria;
ANAS, Azerbaijan;
CNPq and FAPESP, Brazil;
NSERC, NRC and CFI, Canada;
CERN;
ANID, Chile;
CAS, MOST and NSFC, China;
Minciencias, Colombia;
MEYS CR, Czech Republic;
DNRF and DNSRC, Denmark;
IN2P3-CNRS and CEA-DRF/IRFU, France;
SRNSFG, Georgia;
BMBF, HGF and MPG, Germany;
GSRI, Greece;
RGC and Hong Kong SAR, China;
ISF and Benoziyo Center, Israel;
INFN, Italy;
MEXT and JSPS, Japan;
CNRST, Morocco;
NWO, Netherlands;
RCN, Norway;
MEiN, Poland;
FCT, Portugal;
MNE/IFA, Romania;
MESTD, Serbia;
MSSR, Slovakia;
ARRS and MIZ\v{S}, Slovenia;
DSI/NRF, South Africa;
MICINN, Spain;
SRC and Wallenberg Foundation, Sweden;
SERI, SNSF and Cantons of Bern and Geneva, Switzerland;
MOST, Taiwan;
TENMAK, T\"urkiye;
STFC, United Kingdom;
DOE and NSF, United States of America.
In addition, individual groups and members have received support from
BCKDF, CANARIE, Compute Canada and CRC, Canada;
PRIMUS 21/SCI/017 and UNCE SCI/013, Czech Republic;
COST, ERC, ERDF, Horizon 2020 and Marie Sk{\l}odowska-Curie Actions, European Union;
Investissements d'Avenir Labex, Investissements d'Avenir Idex and ANR, France;
DFG and AvH Foundation, Germany;
Herakleitos, Thales and Aristeia programmes co-financed by EU-ESF and the Greek NSRF, Greece;
BSF-NSF and MINERVA, Israel;
Norwegian Financial Mechanism 2014-2021, Norway;
NCN and NAWA, Poland;
La Caixa Banking Foundation, CERCA Programme Generalitat de Catalunya and PROMETEO and GenT Programmes Generalitat Valenciana, Spain;
G\"{o}ran Gustafssons Stiftelse, Sweden;
The Royal Society and Leverhulme Trust, United Kingdom.
 
The crucial computing support from all WLCG partners is acknowledged gratefully, in particular from CERN, the ATLAS Tier-1 facilities at TRIUMF (Canada), NDGF (Denmark, Norway, Sweden), CC-IN2P3 (France), KIT/GridKA (Germany), INFN-CNAF (Italy), NL-T1 (Netherlands), PIC (Spain), ASGC (Taiwan), RAL (UK) and BNL (USA), the Tier-2 facilities worldwide and large non-WLCG resource providers. Major contributors of computing resources are listed in Ref.~\cite{ATL-SOFT-PUB-2021-003}.


\clearpage
\appendix
\part*{Appendix}
\addcontentsline{toc}{part}{Appendix}

For the vector-boson fusion analysis a cut-based selection was also developed, to facilitate reinterpretations of our results outside the ATLAS Collaboration and to serve in parallel to confirm the stability of the results obtained with the ANN.
 
\begin{figure}[!htbp]
\begin{center}
\includegraphics[width=.5\textwidth]{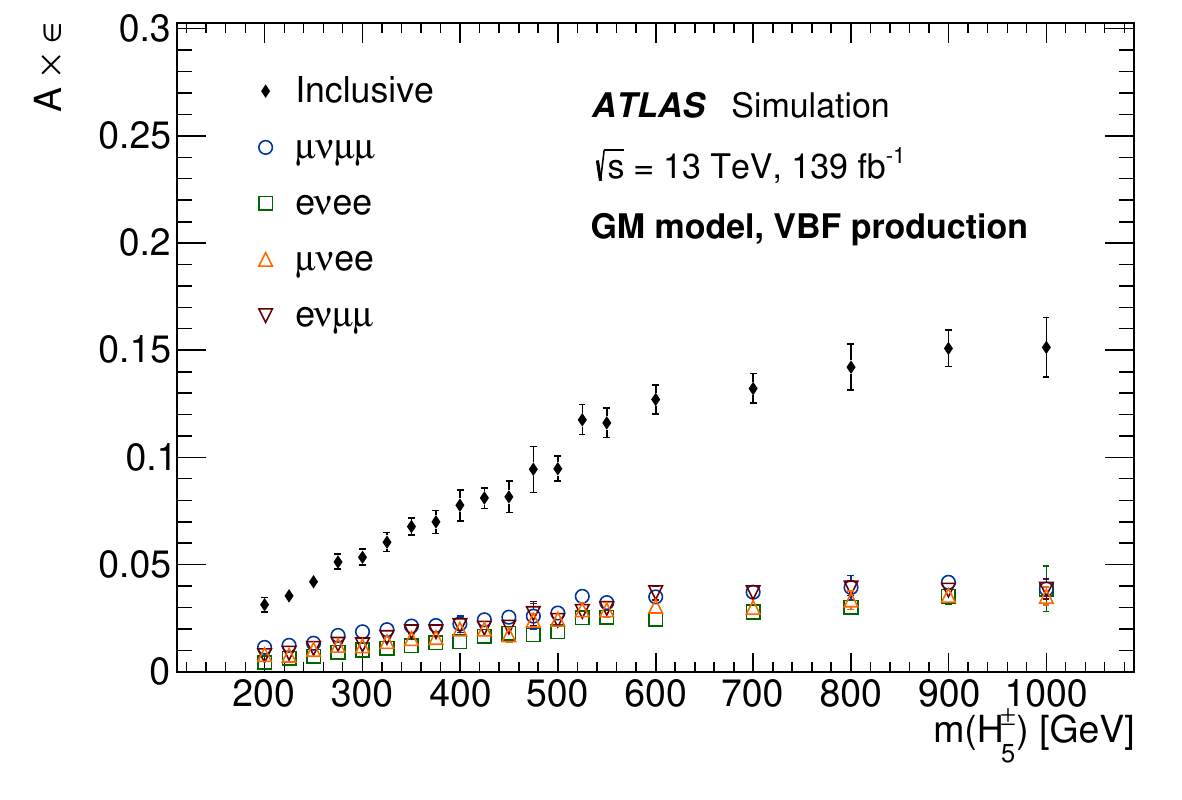}
\includegraphics[width=.5\textwidth]{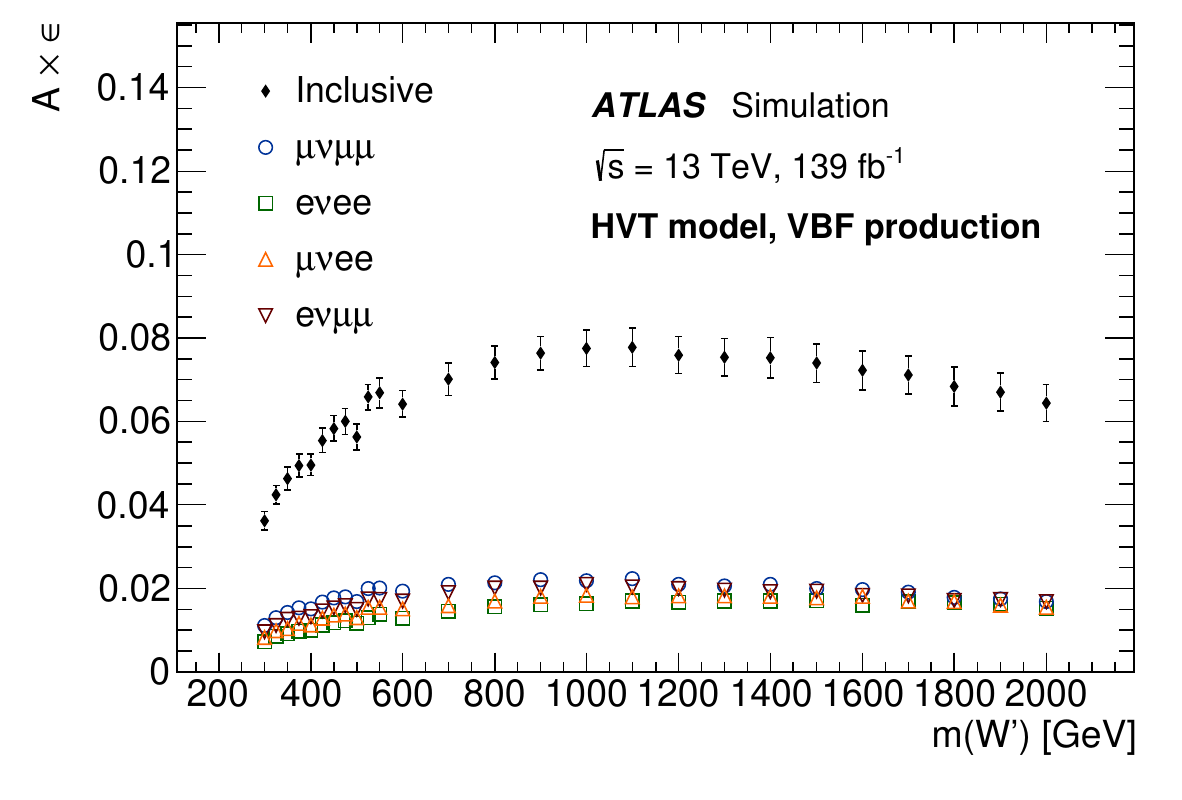}
 
\end{center}
\caption{The acceptance times efficiency of VBF  $H^{\pm}_{5}$ and HVT $W'$ selection using a cut-based VBF selection is shown at different mass points for the individual channels  $\mu^{+}\mu^{-}\mu^{\pm}, \mu^{+}\mu^{-} e^{\pm}, e^{+}e^{-}\mu^{\pm}, e^{+}e^{-}e^{\pm}$ and the sum of all channels.  The error includes statistical and experimental systematic uncertainties.
}
\label{fig:Results:CutBasedAcceptance}
\end{figure}

The VBF cut-based signal region selection is built on top of the baseline \textit{ WZ} selection described in Section~\ref{subsec:preselection} by requiring events to have at least two \textit{VBF jets} with $m_{\textit{jj}}>$500~\GeV~and rapidity separation $|\Delta y_{\textit{jj}}|>$~3.5.
The combined detector acceptance and signal selection efficiency $\cal{A\times\epsilon}$ of the cut-based VBF selection as a function of the mass of the VBF $H^{\pm}_{5}$ and of the HVT $W'$ boson, relative to the generated signal events, is shown in Figure~\ref{fig:Results:CutBasedAcceptance}.
 
\begin{figure}[htbp]
\centering
\subfloat[]{
\includegraphics[width= 0.4\textwidth]{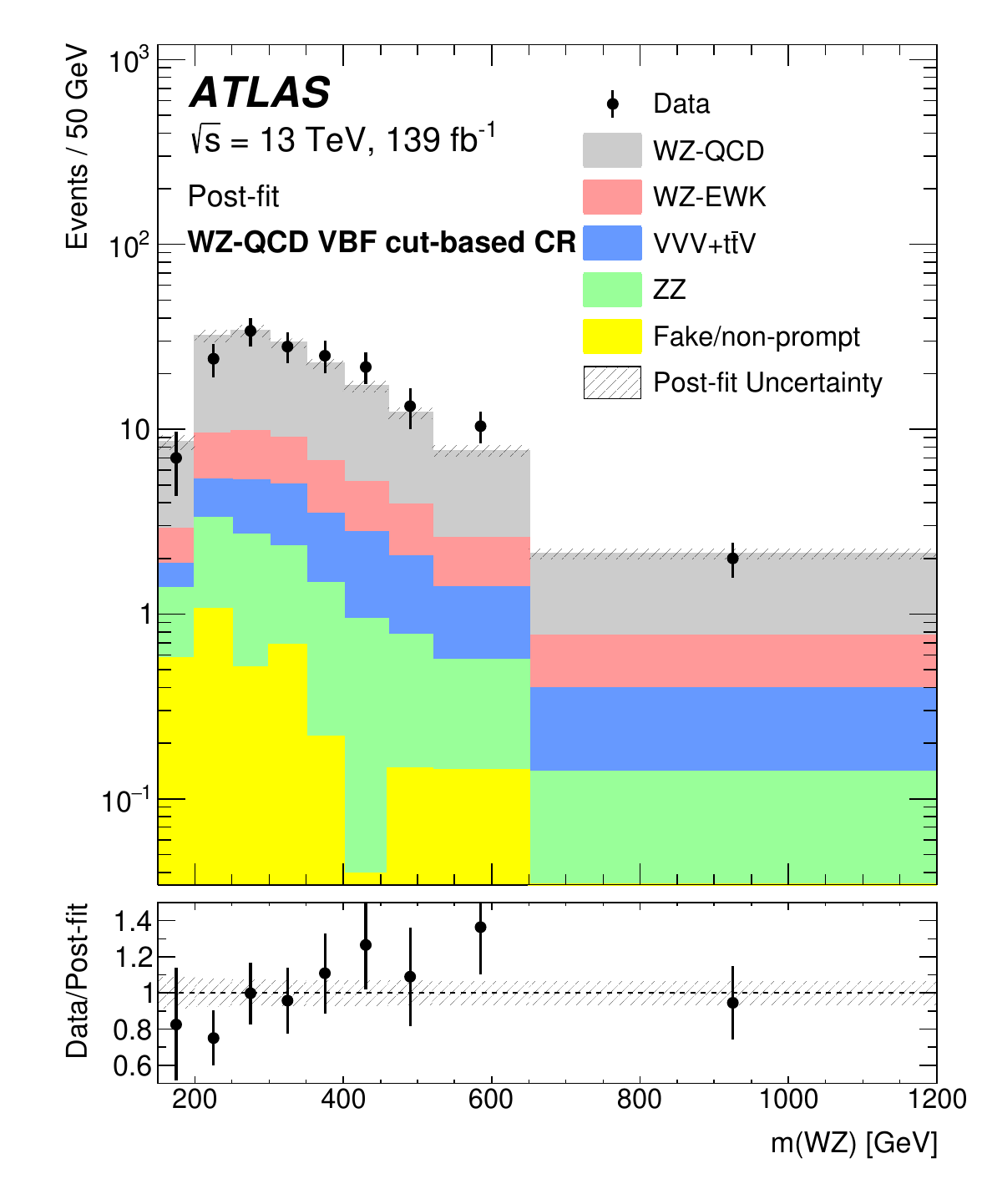}
\label{fig:Results:PostfitCBvbf:WZCR}
}
\subfloat[]{
\includegraphics[width= 0.4\textwidth]{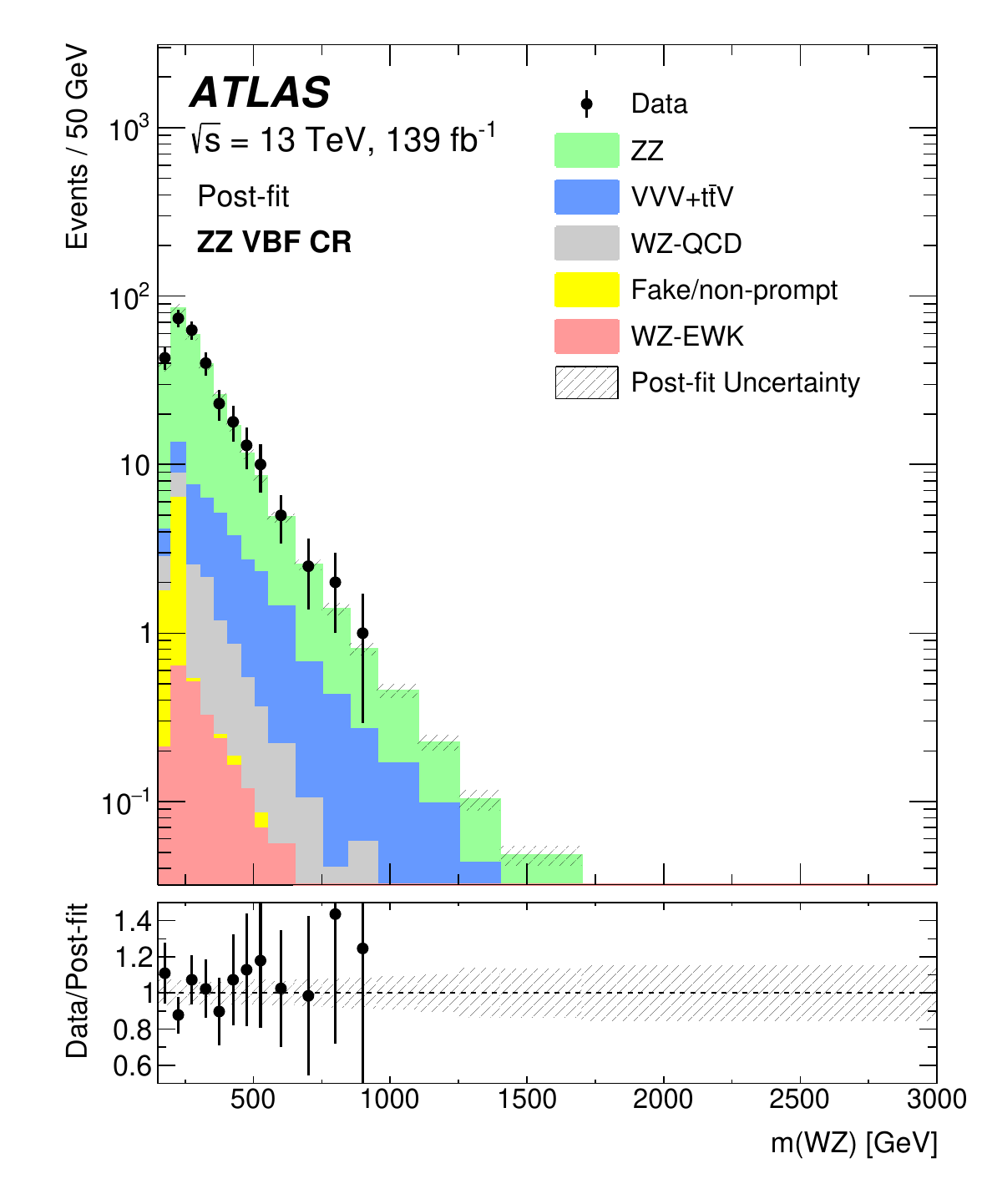}
\label{fig:Results:PostfitCBvbf:ZZCRJJ}
}\\
\subfloat[]{
\includegraphics[width=0.4\textwidth]{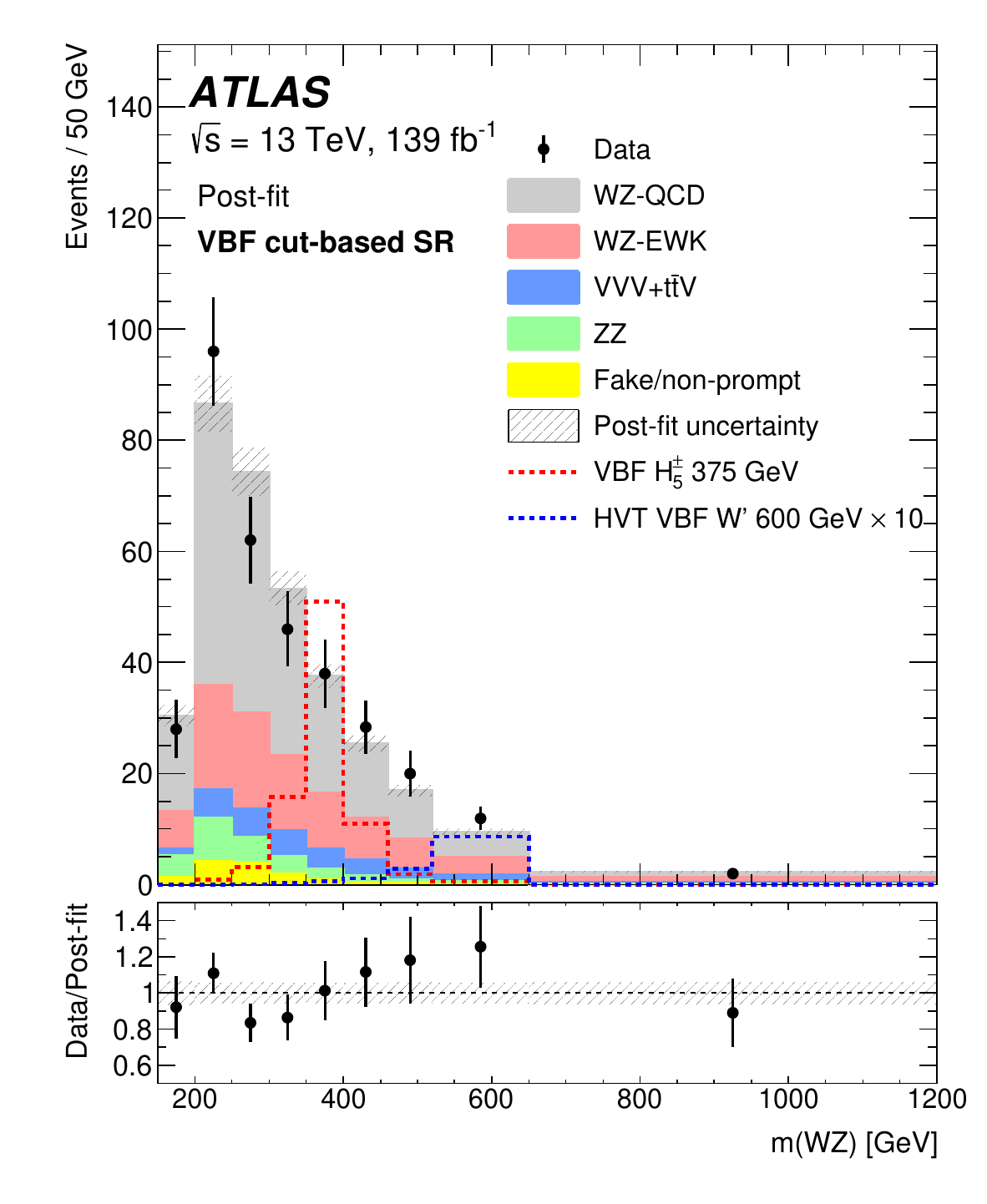}
\label{fig:Results:PostfitCBvbf:SR}
}
\caption{Comparisons of the observed data and the expected background distributions of the $WZ$ invariant mass using the cut based VBF selection are shown for (a) \textit{ WZ}-QCD VBF control region, (b) \textit{ ZZ} VBF control region and (c) the VBF signal region. The background predictions are obtained from a background-only simultaneous fit to the VBF cut-based signal region and the \textit{ WZ}-QCD and \textit{ ZZ} VBF control regions.
For illustration, the expected distributions from an $H_5^{\pm}$ GM model resonance ($\sin \theta_H = 0.5)$ with mass of 375~\GeV~and from an HVT $W'$ (model A) of mass 600~\GeV~are shown in the signal region, with the latter predicted cross sections scaled by 10.
The bottom panel show the ratio of the observed data to the background predictions. The uncertainty in the total background prediction, shown as a hashed area, combines statistical and systematic contributions. }
\label{fig:Results:VBFcutbasedPostfit}
\end{figure}

Following the same strategy as the nominal analysis, two control regions are defined so as to constrain the \textit{ WZ}-QCD and \textit{ ZZ} background. The \textit{ WZ}-QCD VBF cut-based control region is defined by inverting the $|\Delta y_{\textit{jj}}|$ requirement, and the nominal \textit{ ZZ} VBF control region is used. Figure~\ref{fig:Results:VBFcutbasedPostfit} shows comparisons of the data and the expected background distributions. The background predictions are obtained from a background-only simultaneous fit to the VBF cut-based signal region and the \textit{ WZ}-QCD and \textit{ ZZ} VBF control regions.
 
\begin{figure}[!htbp]
\begin{center}
\includegraphics[width=.7\textwidth]{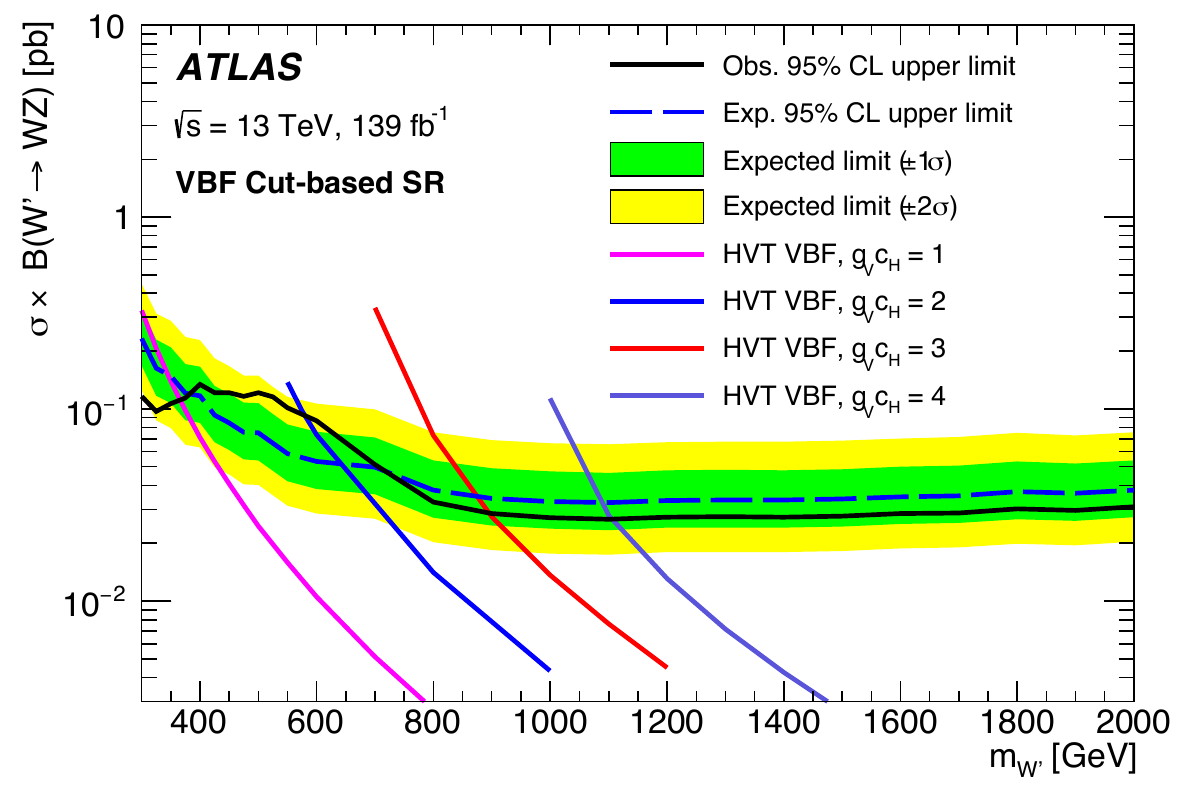}
\end{center}
\caption{Using the cut-based VBF selection, the
observed and expected 95\% CL upper limits  on $\sigma \times B(W' \rightarrow{WZ})$ for the VBF production of a $W'$ boson in the HVT with parameter $c_F$~=~0, as a function of its mass. The LO theory predictions for HVT VBF model with different values of the coupling parameters $g_V$ and $c_H$ are also shown. }
\label{fig:Results:HVTCBLimits}
\end{figure}

Constraints on the production of heavy resonances are derived by repeating the fit to the signal-plus-background hypothesis for different signal models using the cut-based analysis. For the HVT model search, Figure~\ref{fig:Results:HVTCBLimits} presents the observed and expected limits on $\sigma \times B(W' \rightarrow{WZ})$ at 95\%~CL as a function of the $W'$ mass.  Masses below 370~$\GeV$, 590~$\GeV$, 895~$\GeV$ and 1100~$\GeV$ for HVT VBF production can be excluded for the HVT VBF model with $c_F = 0$ and $g_V c_H = 1.0$, 2.0, 3.0 and 4.0, respectively. The sensitivity of the ANN-based analysis is better than that of the cut-based analysis thanks to the utilization of the neural network that better exploits the difference between signals and backgrounds. In the cut-based analysis the relative contribution of background is more important in the signal region, leading to overall higher expected and observed limits. In the region around 375 GeV in particular due to statistics fluctuations in the data both methods of analysis yield consistent observed limits.

\begin{figure}[htbp]
\centering
\subfloat[]{
\includegraphics[width=0.7\textwidth]{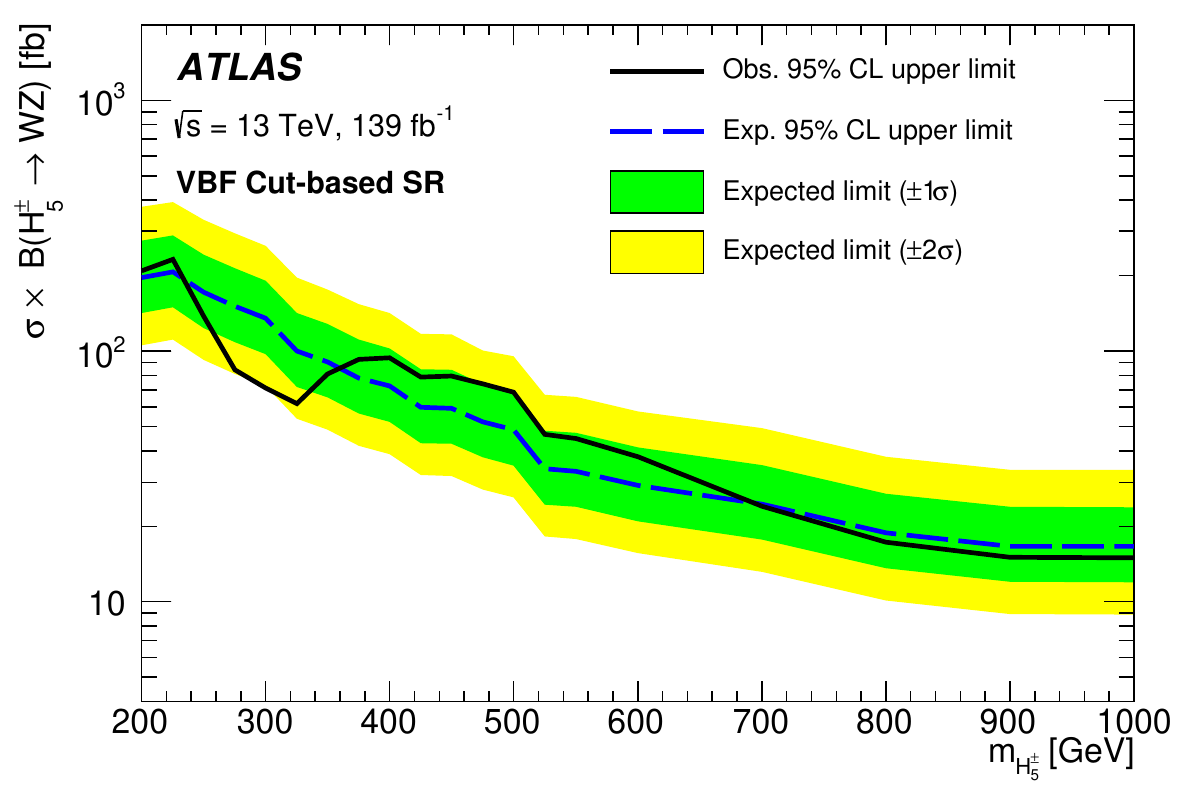}
\label{fig:Results:GMCBLimits:a}
}\\
\subfloat[]{
\includegraphics[width= 0.7\textwidth]{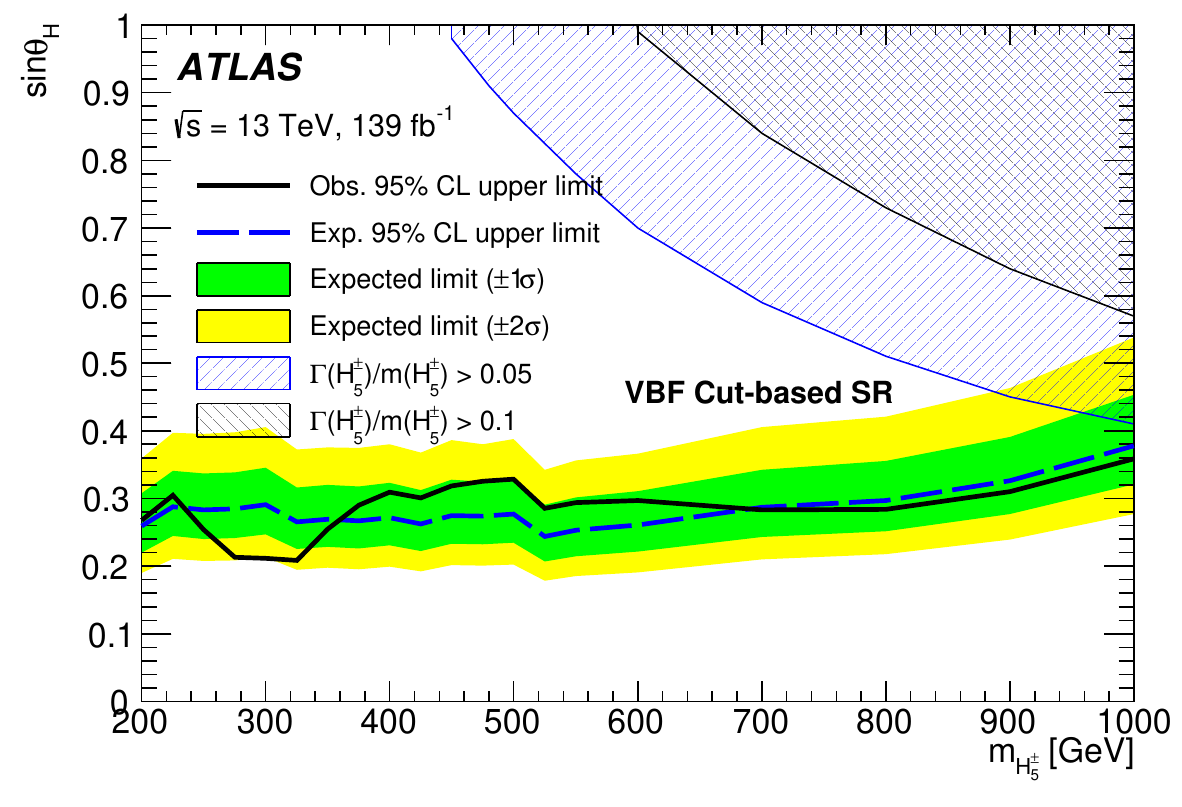}
\label{fig:Results:GMCBLimits:b}
}
\caption{Using the cut-based VBF selection, the observed and expected 95\% CL upper limits on \protect\subref{fig:Results:GMCBLimits:a} the $\sigma \times B(H^{\pm}_5 \rightarrow{WZ})$  and \protect\subref{fig:Results:GMCBLimits:b} the parameter $\sin \theta_H$  of the GM Model as a function of $m_{H^{\pm}_5}$. The shaded regions show where the theoretical intrinsic width of the resonance would be larger than 5\% and 10\% of the mass.}
\label{fig:Results:GMCBLimits}
\end{figure}

For the $H^{\pm}_5$ GM search, observed and expected exclusion limits at 95\% CL on $\sigma \times B(H^{\pm}_5 \rightarrow{WZ})$  and on the mixing parameter $\sin\theta_H$ are shown in Figure~\ref{fig:Results:GMCBLimits}.  The intrinsic width of the scalar resonance, for $\sin\theta_H$=~0.5, is narrower than the detector resolution in the mass region explored. The shaded regions show the parameter space where the $H^{\pm}_5$  width exceeds 5\% and 10\% of $m_{H^{\pm}_5}$. The expected limits extracted using the cut-based analysis are, for both models, between 30\% and 50\% less stringent than the ones extracted using the ANN signal region selection.
 
\clearpage

\printbibliography
 
\clearpage
 
\begin{flushleft}
\hypersetup{urlcolor=black}
{\Large The ATLAS Collaboration}

\bigskip

\AtlasOrcid[0000-0002-6665-4934]{G.~Aad}$^\textrm{\scriptsize 101}$,
\AtlasOrcid[0000-0002-5888-2734]{B.~Abbott}$^\textrm{\scriptsize 119}$,
\AtlasOrcid[0000-0002-7248-3203]{D.C.~Abbott}$^\textrm{\scriptsize 102}$,
\AtlasOrcid[0000-0002-1002-1652]{K.~Abeling}$^\textrm{\scriptsize 55}$,
\AtlasOrcid[0000-0002-8496-9294]{S.H.~Abidi}$^\textrm{\scriptsize 29}$,
\AtlasOrcid[0000-0002-9987-2292]{A.~Aboulhorma}$^\textrm{\scriptsize 35e}$,
\AtlasOrcid[0000-0001-5329-6640]{H.~Abramowicz}$^\textrm{\scriptsize 150}$,
\AtlasOrcid[0000-0002-1599-2896]{H.~Abreu}$^\textrm{\scriptsize 149}$,
\AtlasOrcid[0000-0003-0403-3697]{Y.~Abulaiti}$^\textrm{\scriptsize 116}$,
\AtlasOrcid[0000-0003-0762-7204]{A.C.~Abusleme~Hoffman}$^\textrm{\scriptsize 136a}$
% ...
\bigskip
\\

% ...
$^{29}$Physics Department, Brookhaven National Laboratory, Upton NY; United States of America.\\
% ...
$^{35}$$^{(a)}$Facult\'e des Sciences Ain Chock, R\'eseau Universitaire de Physique des Hautes Energies - Universit\'e Hassan II, Casablanca;$^{(b)}$Facult\'{e} des Sciences, Universit\'{e} Ibn-Tofail, K\'{e}nitra;$^{(c)}$Facult\'e des Sciences Semlalia, Universit\'e Cadi Ayyad, LPHEA-Marrakech;$^{(d)}$LPMR, Facult\'e des Sciences, Universit\'e Mohamed Premier, Oujda;$^{(e)}$Facult\'e des sciences, Universit\'e Mohammed V, Rabat;$^{(f)}$Institute of Applied Physics, Mohammed VI Polytechnic University, Ben Guerir; Morocco.\\
% ...
$^{55}$II. Physikalisches Institut, Georg-August-Universit\"{a}t G\"ottingen, G\"ottingen; Germany.\\
% ...
$^{101}$CPPM, Aix-Marseille Universit\'e, CNRS/IN2P3, Marseille; France.\\
$^{102}$Department of Physics, University of Massachusetts, Amherst MA; United States of America.\\
% ...
$^{116}$Department of Physics, New York University, New York NY; United States of America.\\
% ...
$^{119}$Homer L. Dodge Department of Physics and Astronomy, University of Oklahoma, Norman OK; United States of America.\\
% ...
$^{136}$$^{(a)}$Departamento de F\'isica, Pontificia Universidad Cat\'olica de Chile, Santiago;$^{(b)}$Millennium Institute for Subatomic physics at high energy frontier (SAPHIR), Santiago;$^{(c)}$Instituto de Investigaci\'on Multidisciplinario en Ciencia y Tecnolog\'ia, y Departamento de F\'isica, Universidad de La Serena;$^{(d)}$Universidad Andres Bello, Department of Physics, Santiago;$^{(e)}$Instituto de Alta Investigaci\'on, Universidad de Tarapac\'a, Arica;$^{(f)}$Departamento de F\'isica, Universidad T\'ecnica Federico Santa Mar\'ia, Valpara\'iso; Chile.\\
% ...
$^{149}$Department of Physics, Technion, Israel Institute of Technology, Haifa; Israel.\\
$^{150}$Raymond and Beverly Sackler School of Physics and Astronomy, Tel Aviv University, Tel Aviv; Israel.\\
% ...
$^{a}$ Also Affiliated with an institute covered by a cooperation agreement with CERN.\\
$^{b}$ Also at Borough of Manhattan Community College, City University of New York, New York NY; United States of America.\\
$^{c}$ Also at Bruno Kessler Foundation, Trento; Italy.\\
$^{d}$ Also at Center for High Energy Physics, Peking University; China.\\
$^{e}$ Also at Centro Studi e Ricerche Enrico Fermi; Italy.\\
$^{f}$ Also at CERN, Geneva; Switzerland.\\
% ...

\end{flushleft}


\end{document}